\documentclass[a4paper, fleqn, preprint]{cas-sc}

\usepackage{cleveref}
\usepackage{textcomp}
\usepackage{makecell}
\usepackage[numbers, sort&compress]{natbib}
\usepackage{bm} 

\def\tsc#1{\csdef{#1}{\textsc{\lowercase{#1}}\xspace}}
\tsc{WGM}
\tsc{QE}

\def\R{\mathbb{R}}

\newdefinition{rmk}{Remark}
\newproof{pf}{Proof}
\newproof{pot}{Proof of Theorem \ref{thm}}

\begin{document}

\let\WriteBookmarks\relax
\def\floatpagepagefraction{1}
\def\textpagefraction{.001}

\shorttitle {A geometry-aligned multi-fidelity framework for uncertainty quantification of wildfire spread}

\shortauthors{Vogiatzoglou, et al.,}   

\title [mode = title]{A geometry-aligned multi-fidelity framework for uncertainty quantification of wildfire spread}  

\author[1]{K. Vogiatzoglou}\cormark[1]

\ead{kvogiatzoglou@uth.gr}

\credit{Conceptualization, Data curation, Formal analysis, Software, Validation, Visualization, Writing – original draft}

\author[1]{C. Papadimitriou}

\credit{Supervision, Writing – review and editing}

\author[1]{V. Bontozoglou}

\credit{Supervision, Writing – review and editing}

\affiliation[1]
            {organization={Department of Mechanical Engineering, University of Thessaly},
            address={Leoforos Athinon, Pedion Areos,},
            city={Volos, 38334},
            country={Greece}}

\author[2]{P. Koumoutsakos}

\credit{Supervision, Writing – review and editing}

\affiliation[2]
            {organization={Computational Science and Engineering Laboratory, Harvard University},
            address={19 Oxford St,},
            city={Cambridge, MA 02138},
            country={United States}}

\author[3]{H. Gao}\cormark[1] 

\cortext[cor1]{Corresponding authors}

\ead{hgao1@iastate.edu}

\credit{Conceptualization, Methodology, Project administration, Supervision, Writing – review and editing}

\affiliation[3]
            {organization={Department of Aerospace Engineering, Iowa State University},
            address={537 Bissell Road,},
            city={Ames, IA 50011},
            country={United States}}

\begin{abstract}
Forward propagation of input uncertainties in physics-based wildfire models is computationally prohibitive, limiting the use of high-fidelity simulators in risk assessment workflows. This work introduces a geometry-aligned bi-fidelity surrogate framework that addresses the convection-dominated nature of wildfire spread by mapping low- and high-fidelity solution snapshots onto a common reference domain prior to basis selection and reconstruction. Unlike conventional bi-fidelity schemes, which combine spatially shifted snapshots and consequently suffer from oscillations and excess basis requirements near sharp fronts, the proposed mapping aligns the dominant front geometry through per-variable shift/stretch transforms in 1D and an activity indicator-based affine alignment in 2D, so that reduced bases compare physically corresponding structures rather than displaced ones. Building on the  \textsc{ADfiRe} physics-based simulator, we demonstrate the method on 1D and 2D test cases in which low- and high-fidelity models differ in mesh resolution and physical completeness. Across both settings, the geometry-aligned surrogate reproduces full-field temperature and fuel composition with substantially lower error than its unmapped counterpart, eliminates Gibbs-type oscillations near steep gradients, and recovers high-fidelity probability density functions for key quantities of interest (e.g., maximum temperature, evaporated moisture, and burned area). After offline training, online predictions are roughly three orders of magnitude cheaper than direct high-fidelity evaluation, making the framework a practical building block for many-query uncertainty quantification once the offline cost is amortized over enough queries. We discuss the conditions under which the geometric alignment is most effective, its limitations for non-convex or topologically complex fronts, and the path toward validation against real data.
\end{abstract}

\begin{graphicalabstract}
    \includegraphics[width=1.0\textwidth]{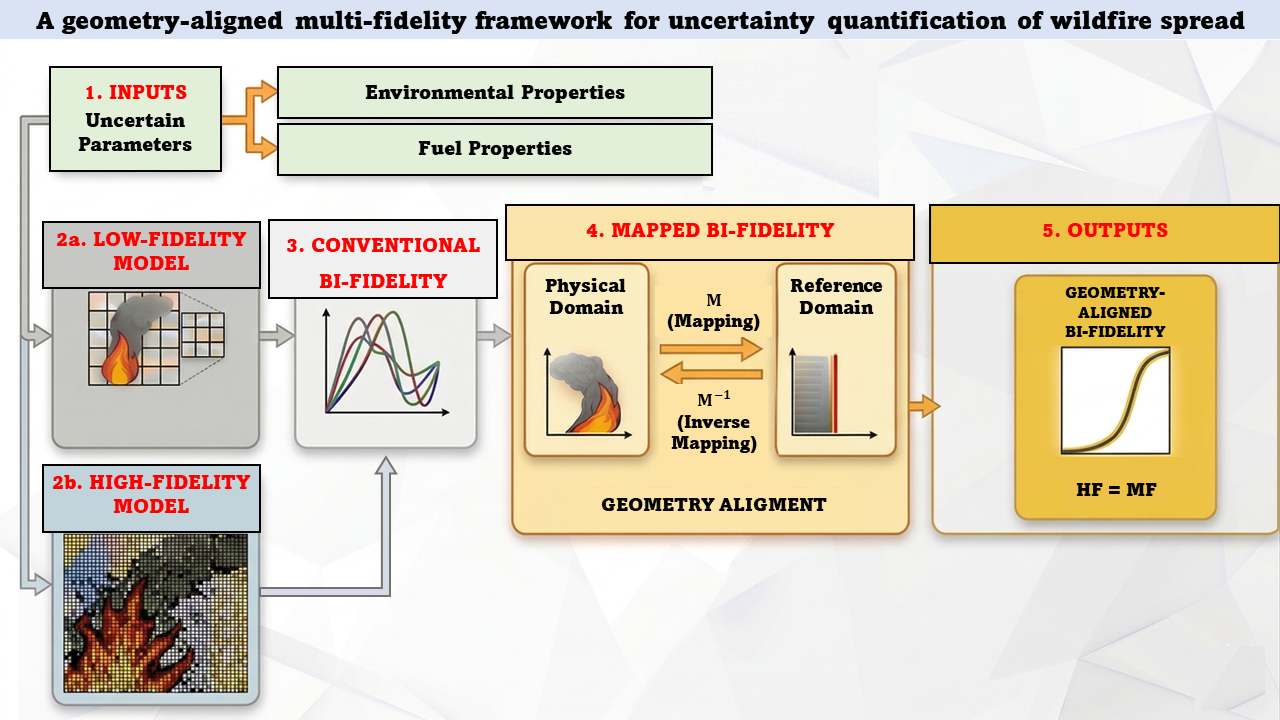}
\end{graphicalabstract}

\begin{highlights}
    \item {A geometry-aligned bi-fidelity method for wildfire spread prediction is introduced.} 
    
    \item {A bijective mapping aligns solution manifolds in convection-dominated regimes.}
    
    \item {Bi-fidelity fields are reconstructed from many low- and few high-fidelity runs.}
    
    \item {Online predictions are three orders of magnitude faster than high-fidelity runs.}
\end{highlights}


\maketitle

\section{Introduction}
    \label{sec:Introduction}

Wildland fires are among the most severe natural hazards worldwide, with intensity and dynamic complexity that have grown markedly in recent decades. Their strong spatiotemporal variability, driven by continuous interactions among fuel characteristics, topography, and the atmospheric boundary layer, underpins what is commonly referred to as extreme fire behavior \citep{Tedim_2018}. Climate change and human pressures have further increased the likelihood of extreme regional fire years across global forested regions \citep{Abatzoglou_2025}, while in Europe the spatial expansion of fire-conducive weather, the shift in seasonal occurrence, and abrupt regime changes increasingly challenge existing adaptive capacity \citep{Giannaros_2023}. These trends motivate the development of predictive tools that are both physically faithful and fast enough to support risk assessment under uncertainty.

Wildfire spread modeling approaches are commonly grouped into three categories: empirical \citep{Rothermel_1972, Fernandes_2009}, physics-based \citep{Mell_2007, Morvan_2004, Vogiatzoglou_2024}, and computational fluid dynamics (CFD) \citep{Linn_2002, Morvan_2011, Finney_2015, Finney_2022}. Empirical models are calibrated to specific experimental conditions and generalize poorly outside them. CFD models offer high physical fidelity but are computationally prohibitive for many-query analyses, restricting their use to research settings. Physics-based models occupy an intermediate position, balancing physical rigor and computational tractability by representing the dominant fire–environment interactions explicitly. Their predictive capability is nonetheless hindered by uncertainty of both aleatoric (e.g., local atmospheric instability, buoyancy-driven gusts) and epistemic (e.g., model parameters, simplified combustion chemistry) origin, which must be quantified rather than ignored.

Partial differential equation (PDE)-based formulations are widely used to model dynamical systems in engineering and applied sciences \citep{Meirovitch_1980, Morvan_2004, Zheng_2009, Vogiatzoglou_2024, Quarteroni_2017,Koumoutsakos_2025}, providing a mechanistic description of transport, reaction, and multiphysics phenomena. Building on this foundation, physics-informed learning approaches integrate data-driven techniques with governing equations, enabling forward predictions that respect physical laws and inverse analyses for parameter identification and uncertainty quantification (UQ). Such approaches, implemented either through deep learning \citep{Lagaris_1998, Raissi_2019, Vogiatzoglou_2025, Apostolakis_2024} or through conservative numerical discretization schemes \citep{Karnakov_2024, Karnakov_2023}, support efficient inference even when observational data are noisy or sparse \citep{Amoudruz_2026}.

UQ is gaining  attention in wildfire modeling \citep{Benali_2016, Riley_2016}, given the substantial uncertainty associated with weather, vegetation, and topography. The central computational difficulty is the forward propagation of input uncertainties to predictions, which typically requires many repeated simulations and quickly becomes prohibitive for physics- and CFD-based models that couple combustion with field-scale and atmospheric dynamics. Recent work has begun to mitigate this cost through multi-fidelity strategies. Schwerdtner et al. \citep{Schwerdtner_2024} showed that surrogates trained on correlated low-cost datasets enable scalable multi-fidelity UQ for large-scale wildfire simulations, reducing training times from months to hours while improving accuracy. Similarly, Valero et al. \citep{Valero_2021} combined high-fidelity CFD with coarser-resolution and semi-empirical models through Monte Carlo (MC)-based variance-reduction techniques, achieving substantial speedups over standard MC while preserving fire-spread accuracy. Complementary efforts have used probabilistic frameworks for meteorological and fuel-structure inputs \citep{Anderson_2007, Cruz_2017}, Taylor-series–based methods \citep{Bachmann_2002}, and acceleration techniques for classical MC schemes built on Rothermel's model \citep{Jimenez_2008} and on empirical heat-transfer formulations \citep{Yuan_2020}.

Despite these advances, a methodological gap remains. Wildfire spread is dominated by sharp, advecting combustion fronts, and the position of these fronts varies strongly with the uncertain inputs. Conventional bi-fidelity surrogates \citep{Zhu_2014, Gao_2020, Gao_2021} that combine low-fidelity (LF) and high-fidelity (HF) snapshots through linear projections in the physical domain therefore compare spatially shifted features, leading to slow basis convergence (i.e., slow Kolmogorov $N$-width decay \citep{GreifUrban_2019, Arbes_2025}), Gibbs-type oscillations \citep{Gottlieb_1997} near steep gradients, and non-physical smoothing of the front. These pathologies are well documented for convection-dominated reduced-order models \citep{Ohlberger_2016, Mirhoseini_2023}, where domain mappings that align parameter-dependent features within a common reference frame have been shown to recover an effectively low-dimensional representation by rendering the dominant convective structures stationary in the transformed space.

In this work, we exploit this observation in a multi-fidelity setting. Building on the existing physics-based simulator \textsc{ADfiRe} \citep{Vogiatzoglou_2024}, whose full-scale formulation we adopt as the HF model and whose simplified configurations serve as LF counterparts, we introduce a geometry-aligned bi-fidelity surrogate framework for UQ in wildfire spread. The proposed model is formulated as an advection–diffusion–reaction (ADR) system in which fire-front propagation evolves spatiotemporally as a function of uncertain inputs. The key methodological contribution is a bijective mapping that transports LF and HF solution snapshots into a common reference domain prior to basis selection and reconstruction, so that reduced bases compare physically corresponding structures rather than displaced ones. Two variants of this mapping are developed: a per-variable shift/stretch transform tailored to the one-dimensional (1D) setting and an activity indicator–based affine alignment for the two-dimensional (2D) setting. Front geometry and field reconstruction are thereby decoupled, with front geometry described by a small set of geometric descriptors and the reconstructed field returned to the physical domain through the inverse mapping.

The proposed framework offers three concrete advantages relative to the corresponding conventional bi-fidelity surrogate. First, it eliminates the spurious oscillations introduced by combining shifted snapshots and significantly improves the reconstruction accuracy of full-field temperature and fuel composition (convection-dominated dynamics). Second, it recovers HF probability density functions of physically meaningful quantities of interest, including maximum temperature, total evaporated moisture, and total burned area, from a small ensemble of HF evaluations. Third, once the offline phase is completed, online predictions are roughly three orders of magnitude cheaper than direct HF simulation, making the framework a practical building block for many-query UQ workflows after the offline cost has been amortized over a sufficient number of queries. We also discuss the regimes in which the alignment is most effective, in particular for sharply localized convection-dominated fronts, and the limitations that arise when fronts become topologically complex or non-convex.

The remainder of this paper is organized as follows. Section~\ref{sec:Methodology} describes the \textsc{ADfiRe} physics-based simulator and the proposed methodology, including the offline training and online prediction phases as well as the geometry-mapping construction. Section~\ref{sec:Results} reports results for 1D and 2D simulated case studies. Section~\ref{sec:Conclusion} summarizes the main findings, discusses limitations, and outlines directions for future work, including validation against real wildfire events. 

\section{Methodology}
    \label{sec:Methodology} 

This section presents the two fundamental components of this work: (i) the \textsc{ADfiRe} physics-based wildfire simulator and (ii) the proposed geometry-aligned bi-fidelity methodology.

\subsection{Physics-based wildfire spread model}
    \label{sec:Model}

Effective wildfire preparedness relies on simulation tools that balance computational efficiency with physical realism, while accommodating spatially heterogeneous fuel properties, weather conditions, and terrain features. To this end, we previously developed a physics- and data-informed wildfire spread modeling framework \citep{Vogiatzoglou_2024} that couples mechanistic fire dynamics, formulated through an ADR system, with empirical information \citep{Andrews_2018, Rothermel_1972, Scott_2005, Fernandes_2009} and insights from high-fidelity CFD simulations \citep{Linn_2002, Morvan_2011, Finney_2015, Finney_2022}. This framework constitutes the core of the \textsc{ADfiRe} simulator, which is designed for efficient wildfire-spread analysis in complex fuel--weather--environment settings. The underlying physical formulation is summarized below.

Wildfire modeling requires the integration of multiple spatiotemporal scales and interacting processes, including fuel combustion chemistry and the governing physics of fluid flow and heat--mass transport. Within this approach, the energy balance incorporates both endothermic and exothermic source terms, representing water evaporation and fuel combustion, respectively. Heat transfer within the control volume accounts for advection together with diffusion-like, short-range interactions. The latter encompass turbulent and buoyant transport, wind-induced dispersion, and dense-medium radiative effects. Thermal losses from the upper boundary of the control volume to the ambient environment, arising from natural convection and radiation, are also included. Building upon our previous work \citep{Vogiatzoglou_2024}, the proposed model predicts fire propagation over arbitrary 1D and 2D topography. Consistent with the formulation, the fluid-mechanics problem is not solved explicitly, instead, the flow field required in the energy balance is introduced through heuristic considerations, enhancing computational efficiency.

In accordance with the simplified representation of fuel composition, combustion is modeled as a two-stage sequential process. The initial stage is endothermic and accounts for moisture evaporation and fuel pyrolysis, leading to the formation of volatiles and char (combustibles). This is followed by an exothermic stage associated with the oxidation of combustibles. The endothermic stage is described using first-order Arrhenius kinetics:
\begin{equation}
    \label{eq:Arrhenius_Endothermic}
    r_{e} =c_{e} \, e^{-\frac{b_{e}}{T}}.
\end{equation}
The exothermic stage is likewise modeled using first-order Arrhenius kinetics, with reaction rates dependent on both temperature (kinetics-dominated combustion) and oxygen availability (mass-transfer-dominated combustion), yielding the following expression for the depletion of combustibles:
\begin{equation}
    \label{eq:Arrhenius_Exothermic}
    r_{x} = \frac{c_{x} \, e^{-\frac{b_x}{T}} \, r_{o}}{c_{x} \, e^{-\frac{b_{x}}{T}} + r_{o}},
\end{equation}
where $T$ [K] denotes the fire-layer temperature, $c_{e}, c_{x}$ [s$^{-1}$] together with $b_{e}, b_{x}$ [K] are model parameters characterizing the ignition and combustion processes, and $r_{o}$ [s$^{-1}$] represents the effective oxygen availability during the pyrolysis stage.

The model is formulated through a thermal energy balance that captures the evolution of temperature $T(x,y, t)$ together with the influence of fuel composition, including moisture content $S_{e}(x,y, t)$ and combustible material $S_{x}(x,y, t)$. The burning fuel and the surrounding hot gases are assumed to share a single representative temperature, taken to be uniform in the vertical ($z$) direction while varying spatially in the horizontal plane $(x, y)$ and in time ($t$). By considering a control volume of unit ground area extending vertically up to the maximum canopy height, the resulting wildfire physics-based model (\textsc{ADfiRe}) reads:
\begin{equation}
    \label{eq:Wildfire_Model}
    \begin{aligned}
    \frac{\partial T}{\partial t} &= \kappa_{1} \Big(\boldsymbol{\nabla}\cdot(\mathbf{D} \, \boldsymbol{\nabla} T)
    - \boldsymbol{\nabla}\cdot(\mathbf{u} \, T) \Big)
    - \kappa_{2} \, S_{e} \, r_{e} + \kappa_{3} \, S_{x} \, r_{x}
    - \kappa_{4} U \left(T - T_{a}\right),\\
    \frac{\partial S_{e}}{\partial t} &= - S_{e} \, r_{e},\\
    \frac{\partial S_{x}}{\partial t} &= - S_{x} \, r_{x},
    \end{aligned}
\end{equation}
where $T, S_{e}$, and $S_{x}$ denote the temperature distribution and the endothermic (moisture) and exothermic (combustibles) fuel mass fractions, respectively. The coefficients $\kappa_{1}, \kappa_{2}, \kappa_{3},$ and $\kappa_{4}$ encapsulate fuel, combustion, and environmental properties, $T_{a}$ [K] is the ambient temperature, the diffusion coefficient vector is $\mathbf{D} = (D_{x}, D_{y})$ [m$^{2}$ s$^{-1}$], and the velocity vector $\mathbf{u} = (u_{x}, u_{y})$ [m s$^{-1}$] represents the gas flow averaged over the plantation height. Finally, $U$ [W m$^{-2}$ K$^{-1}$] denotes the overall heat transfer coefficient governing thermal losses to the surrounding environment, and $\boldsymbol{\nabla}$ is the gradient operator. All physical quantities are discussed in detail in our previous work \citep{Vogiatzoglou_2024}. We briefly revisit the diffusion coefficient $\mathbf{D}$ next, as it plays a central role in defining the LF and HF surrogate models in the application section.

For the numerical discretization model, a conservative finite-difference scheme is employed, using upwind differencing for advection and central differences for diffusion to ensure stability in convection-dominated regimes. Open-outflow boundary conditions are imposed to allow the unimpeded transport of heat and species across the domain boundaries.

Owing to the high soot concentrations typical of wildland fires, the fire zone may be treated as an optically dense medium. Under this assumption, radiative heat transfer is approximated using the Rosseland diffusion model \citep{Siegel_1972}, $q_{r}=- \, k_{r} \, \nabla T$, where the equivalent radiative diffusivity is $k_{r}=5.33 \, \sigma_b \, T^3/\left(1-e^{-k_dL}\right)$. Here $\sigma_b$ [W m$^{-2}$ K$^{-4}$] is the Stefan--Boltzmann constant, $k_d$ [m$^{-2}$] is the absorption coefficient, and $L$ [m] is the fire-layer depth. Accounting for buoyancy-induced motion, wind-driven shear, and radiative transport, the diffusivity coefficient is modeled as:
\begin{equation}
    \label{eq:Diffusion_Coefficient}
    \mathbf{D} = D_{b} + A_{d} \, \mathbf{u}\,L\,(1-e^{-\gamma_{d}\, w}) \, + \frac{k_r}{\rho_g\,c_{pg}} \,,
\end{equation}
where $D_{b}$ [m$^{2}$ s$^{-1}$] denotes a representative coefficient associated with buoyancy-driven motion, $A_{d}$ is a dispersion constant, $\rho_{g}$ [kg m$^{-3}$] is the gas density, and $c_{pg}$ [J kg$^{-1}$ K$^{-1}$] is the specific heat capacity of the hot gases. Finally, $L$ and $w$ [m] represent the fireline depth and width, respectively.

\subsection{Bi-fidelity surrogate structure}
    \label{sec:BiFidelity_Structure} 

\subsubsection{Algorithmic overview}
    \label{sec:Algorithmic_Overview}

\begin{figure}
    \centering    
    \includegraphics[width=.8\textwidth]{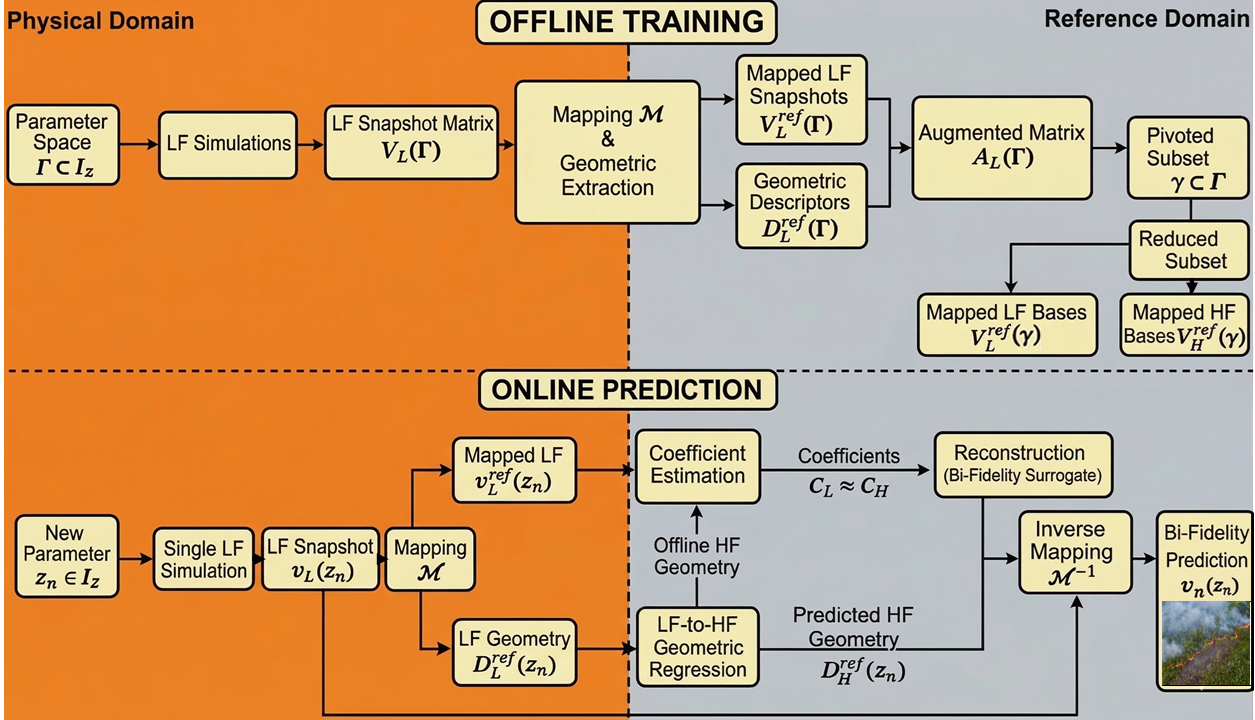}
    \caption{Schematic representation of the geometry-aligned bi-fidelity algorithm, including offline construction and online prediction stages across both the physical and reference domains.}
    \label{fig:Figure_1}
\end{figure}

The proposed bi-fidelity framework constructs a geometry-aligned surrogate predictor that combines abundant LF solution snapshots with a limited number of HF evaluations. Building on the general principles of multi-fidelity stochastic collocation and bi-fidelity approximation~\citep{Narayan_2014, Zhu_2014, Gao_2020, Gao_2021}, the present method introduces a front-aligned mapping strategy tailored to convection-dominated wildfire dynamics, enabling LF and HF solution manifolds to be compared and reconstructed in a common reference domain. The resulting surrogate--an efficient representation of the HF response--is informed primarily by the structure of the LF solutions, with corrections introduced through a small number of strategically selected HF evaluations. The methodology is developed in two phases (see Fig.~\ref{fig:Figure_1}):

\begin{enumerate}
    \item In the offline training phase, a large ensemble of LF simulations is generated by sampling the uncertain parameter space. These LF snapshots are analyzed to identify the most informative collocation points at which HF simulations are then performed. To account for the convection-dominated nature of the problem, both LF and HF solution bases are subsequently mapped onto a common reference domain (geometry-aligned mapping), enabling coherent comparison and projection.

    \item In the online prediction phase, the bi-fidelity surrogate is evaluated at arbitrary collocation points through a projection-based reconstruction that combines the precomputed LF and HF bases from the offline phase. The surrogate solution is assembled using predominantly LF information while retaining HF-level accuracy, yielding substantial computational savings.
\end{enumerate}

\subsubsection{Offline training strategy}
    \label{sec:Offline_Phase}

In the offline (or pre-computation) stage, the objective is to characterize the solution behavior across the admissible parameter domain $I_{z}$. This is achieved by performing a sufficiently large number ($M \gg 1$) of LF simulations, exploiting their computational efficiency while still capturing the dominant physical trends of the system. The uncertain parameter vector $\mathbf{z}$, representing weather and fuel-related properties in our work, is sampled over $I_{z}$ to construct a discrete training dataset,
\begin{equation}
    \label{eq:Set_Gamma}
    \Gamma \, = \, \big\{\mathbf{z_{1}}, \, \mathbf{z_{2}}, \, \ldots, \mathbf{z_{M}}\big\} \subset I_{z}.
\end{equation}
The sampling method may rely on Latin Hypercube Sampling (LHS) \citep{MCKay_1979}, classical Monte Carlo \citep{Ching_2007}, or other efficient strategies such as importance sampling \citep{Rubinstein_1999} or adaptive/sequential sampling \citep{Echard_2011}, provided that the selected points cover the region of interest $I_{z}$. For each sampled realization $\mathbf{z} \in \Gamma$, an LF simulation is performed, producing the corresponding final snapshot $v_{L}(\mathbf{z})$. Collectively, these realizations form the LF snapshot matrix:
\begin{equation}
    \label{eq:LF_Matrix}
    V_{L}(\Gamma) \, = \, \big[v_{L}(\mathbf{z_{1}}), \, v_{L}(\mathbf{z_{2}}), \, \ldots, v_{L}(\mathbf{z_{M}})\big]^{\top}.
\end{equation}
Every snapshot $v_{L}(\mathbf{z})$ contains the discrete solutions of all state variables, namely temperature and fuel components (moisture and combustibles), as shown below:
\begin{equation}
    \label{eq:LF_Snapshot}
    v_{L}(\mathbf{z}) \, = \,
    \begin{bmatrix}
    T_{L}(\mathbf{z}) \\
    S_{e_{L}}(\mathbf{z}) \\
    S_{x_{L}}(\mathbf{z})
    \end{bmatrix}.
\end{equation}
Throughout the methodology, an affine scaling (normalization) based on reference states is applied so that all state variables lie within $[0,1]$, preventing any single variable from disproportionately influencing the subsequent analysis.

The core innovative component of the present work consists in mapping all LF snapshots from the physical (original) domain to a common reference (latent) domain \citep{Mirhoseini_2023}. Owing to the predominantly convection-driven nature of wildfire propagation, each LF snapshot $v_{L}(\mathbf{z})$ exhibits a combustion front with a distinct spatial location and geometric character. To enable meaningful comparison and an efficient reduced-order representation, a bijective transformation from the physical to the reference space, $V_{L}(\Gamma) \, \mapsto \, V_{L}^{ref}(\Gamma)$, is introduced. The construction of this mapping is detailed for both the 1D and 2D formulations in \S\ref{sec:Mapping_Phase}. The geometric descriptors associated with each transformation are computed and retained for subsequent use in the online phase. For example, in the 1D setting these descriptors consist of variable-specific shift and stretch parameters (e.g., $s_{T}, s_{S_e}, s_{S_x}, k_{S_e}, k_{S_x}$), whereas in the 2D formulation the mapping is characterized by the statistical moments of the activity indicator (e.g., $\mu_{x}, \mu_{y}, \sigma_{x}, \sigma_{y}$). For notational convenience, all such geometric descriptors are collectively denoted as $D_{L}^{ref}(\Gamma)$.

Given the nodal set $\Gamma$ and the associated LF snapshot matrix, the next step is to select a reduced subset of $m$ representative parameter realizations,
\begin{equation}
    \label{eq:Set_gamma}
    \gamma\, = \, \big\{\mathbf{z_{1}}, \, \mathbf{z_{2}}, \, \ldots, \mathbf{z_{m}}\big\} \subset \Gamma,
\end{equation}
with $m \ll M$. This reduced subset defines the locations at which the computationally expensive HF simulations are performed, yielding the HF snapshot matrix:
\begin{equation}
    \label{eq:HF_Matrix}
    V_{H}(\gamma) \, = \, \big[v_{H}(\mathbf{z_{1}}), \, v_{H}(\mathbf{z_{2}}), \, \ldots, v_{H}(\mathbf{z_{m}})\big]^{\top}.
\end{equation}
The same bijective transformation is then applied to map the HF snapshots from the physical to the reference space, $V_{H}(\gamma) \, \mapsto \, V_{H}^{ref}(\gamma)$.

To select the representative subset $\gamma \subset \Gamma$ used for the HF evaluations, an LF-informed greedy strategy is adopted \citep{Narayan_2014, Zhu_2014}. The procedure constructs the reduced set sequentially, ensuring that each newly selected parameter point contributes the maximum additional information relative to those previously chosen. Let $\gamma^{(k-1)} \, = \, \{\mathbf{z_{1}}, \, \mathbf{z_{2}}, \, \ldots, \mathbf{z_{k-1}}\}$ denote the subset identified at iteration $k-1$, and let $V_{L}^{ref}(\gamma^{(k-1)}) \, = \, \operatorname{span}\big\{v_{L}^{ref}(\mathbf{z_{1}}), \, v_{L}^{ref}(\mathbf{z_{2}}), \, \ldots, v_{L}^{ref}(\mathbf{z_{k-1}})\big\}$ be the associated LF subspace. The next parameter realization is chosen as the one whose LF snapshot yields the largest projection residual relative to the current subspace:
\begin{equation}
    \label{eq:Greedy_Algorithm}
    \mathbf{z_k} \, = \, \arg\max_{\mathbf{z} \, \in \, \Gamma} \left\|\, v_{L}^{ref}(\mathbf{z}) - \operatorname{proj}_{V_{L}^{ref}(\gamma^{(k-1)})}\, v_{L}^{ref}(\mathbf{z}) \,\right\|,
\end{equation}
where the second term on the right-hand side denotes the orthogonal projection. The subset is updated at every iteration as $\gamma^{(k)} \, = \, \gamma^{(k-1)} \cup \{\mathbf{z_k}\}$. For computational efficiency, this distance-based selection is implemented through factorization of the augmented-LF Gramian matrix $G \in \mathbb{R}^{M \times M}$, defined by,
\begin{equation}
    \label{eq:Gramian_Matrix}
    G_{ij} \, = \, \big\langle A_{L}(\mathbf{z_{i}}), \, A_{L}(\mathbf{z_{j}}) \big\rangle,
\end{equation}
where $\langle \cdot , \cdot \rangle$ denotes the inner product in the augmented-LF solution space and $A_{L}$ is the augmented-LF matrix. A pivoted Cholesky decomposition,
\begin{equation}
    \label{eq:Pivoted_Cholesky}
    G \, = \, P^{\top} \mathcal{L} \mathcal{L}^{\top} P,
\end{equation}
with $\mathcal{L}$ lower triangular and $P$ a permutation matrix, is employed to rank the LF snapshots according to their linear independence. The permutation $P$ provides an ordered sequence of parameter realizations, and the indices corresponding to the first $m$-pivots define the subset $\gamma$ used for the HF simulations. In practice, only the truncated Gramian associated with these leading pivots is required, and its condition number is monitored to ensure numerical stability of the reduced representation. The augmented-LF reference-based snapshot matrix used for node selection is constructed as:
\begin{equation}
    \label{eq:Augmented_Matrix}
    A_{L}(\Gamma) =
    \begin{bmatrix}
    V_{L}^{ref}(\Gamma) \\
    \beta \cdot D_{L}^{ref}(\Gamma)
    \end{bmatrix} \, = \,
    \begin{bmatrix}
    T_{L}^{ref}(\Gamma) \\
    S_{e_{L}}^{ref}(\Gamma) \\
    S_{x_{L}}^{ref}(\Gamma) \\
    \beta \cdot D_{L}^{ref}(\Gamma)
    \end{bmatrix},
\end{equation}
with $\beta$ a weighting parameter. This augmentation ensures that the selection of the $m$-collocation points captures both the intrinsic field dynamics (sampled snapshots) and the associated front geometry (geometric descriptors).

At this stage, the collocation points have been identified and the corresponding LF and HF bases have been constructed for use in the online prediction phase. The reduced approximation spaces spanned by these bases are:
\begin{align}
    \label{eq:LF_HF_Bases}
    U_{L}^{ref}(\gamma) \, &= \, \operatorname{span}\!\left(V_{L}^{ref}(\gamma)\right)
    \, = \, \operatorname{span}\big\{v_{L}^{ref}(\mathbf{z_{1}}), \,  v_{L}^{ref}(\mathbf{z_{2}}), \, \ldots, v_{L}^{ref}(\mathbf{z_{m}})\big\}, \\
    U_{H}^{ref}(\gamma) \,  &= \,  \operatorname{span}\!\left(V_{H}^{ref}(\gamma)\right)
    = \, \operatorname{span}\big\{v_{H}^{ref}(\mathbf{z_{1}}), \,  v_{H}^{ref}(\mathbf{z_{2}}), \, \ldots, v_{H}^{ref}(\mathbf{z_{m}})\big\}.
\end{align}

\subsubsection{Online bi-fidelity surrogate evaluation}
    \label{sec:Online_Phase}

In the online stage, the precomputed LF and HF reduced bases associated with the subset $\gamma$ are exploited to generate predictions for previously unseen parameter instances. These bases define the reduced approximation spaces $U_{L}^{ref}$ and $U_{H}^{ref}$ in the reference domain. The primary advantage is that, for any new parameter realization $\mathbf{z_{n}} \in I_{z}$, only a single LF simulation is required, which reduces the per-query cost by approximately three orders of magnitude in the cases considered here. The corresponding HF response (the bi-fidelity prediction) is recovered through the precomputed reduced representation, without performing an additional HF simulation. Specifically, the LF model is evaluated at a new point $\mathbf{z_{n}}$, yielding the physical-domain snapshot $v_{L}(\mathbf{z_{n}})$. Consistent with the mapping-based formulation, this solution is then transformed into the reference domain, $v_{L}(\mathbf{z_{n}}) \mapsto v_{L}^{ref}(\mathbf{z_{n}})$. The associated HF response is reconstructed in the reference space and mapped back to the physical domain, completing the prediction without requiring a direct HF computation.

This reconstruction relies on the central assumption of the bi-fidelity framework: both LF and HF solutions admit reduced representations within their respective approximation spaces, and the LF and HF expansions share the same coefficient vector \citep{Narayan_2014, Zhu_2014, Gao_2020, Gao_2021}, denoted by $C$. In other words, if the LF solution can be expressed as a linear combination of the LF basis vectors, the corresponding HF solution can be reconstructed by applying the same coefficients to the HF basis vectors. The distinctive aspect of the present work is the per-variable treatment of these coefficients. Given the multiphysics nature of wildfire dynamics, each state variable is treated independently in the reduced approximation. This separation is motivated by the distinct spatiotemporal characteristics of the physical fields: temperature exhibits a localized combustion-front structure with steep gradients near the fireline, whereas the fuel components (moisture and combustibles) exhibit sharper depletion fronts and predominantly decrease over time. Accordingly, separate reduced representations are adopted for each state variable in the LF reference domain:
\begin{align}
    \label{eq:New_LF}
    T_{L,n}^{ref} \, = \, T_{L}^{ref}(\mathbf{z_{n}}) \,  &\approx \, T_{L}^{ref}(\gamma) \, C_{L,T}, \\
    S_{e_{L, n}}^{ref} \, = \, S_{e_{L}}^{ref}(\mathbf{z_{n}}) \, &\approx \, S_{e_{L}}^{ref}(\gamma) \, C_{L,S_{e}}, \\
    S_{x_{L, n}}^{ref} \, = \, S_{x_{L}}^{ref}(\mathbf{z_{n}}) \, &\approx \, S_{x_{L}}^{ref}(\gamma) \, C_{L,S_{x}},
\end{align}
where $C_{L,T}, C_{L,S_{e}}, C_{L,S_{x}}$ denote the expansion coefficients for each state variable. This field-wise decomposition allows distinct reduced subspaces tailored to the dynamics of each variable. Compared to a unified formulation employing a single global coefficient vector, this approach offers improved flexibility and numerical stability. A unified representation was also investigated and resulted in inferior approximation accuracy. The expansion coefficients are obtained by solving a regularized least-squares projection problem for each variable:
\begin{align}
    \label{eq:Projection_Method}
    \arg\min_{C_{L,T}} \; \big\| T_{L}^{ref}(\gamma) \, C_{L,T} - T_{L,n}^{ref} \big\|_2 + \lambda \big\| C_{L,T} \big\|_2, \\
    \arg\min_{C_{L,S_{e}}} \; \big\| S_{e_{L}}^{ref}(\gamma)\, C_{L,S_{e}} - S_{e_{L, n}}^{ref}\big\|_2 + \lambda \big\| C_{L,S_{e}} \big\|_2, \\
    \arg\min_{C_{L,S_{x}}} \; \big\| S_{x_{L}}^{ref}(\gamma)\, C_{L,S_{x}} - S_{x_{L, n}}^{ref}\big\|_2 + \lambda \big\| C_{L,S_{x}} \big\|_2.
\end{align}
Here $\lambda > 0$ is the Tikhonov regularization parameter, which penalizes excessively large coefficients, enhances stability under ill-conditioning, and mitigates linear dependencies within the reduced basis. Assuming coefficient consistency between LF and HF representations, $C_{L,T} \approx C_{H,T} \equiv C_{T}$, $C_{L,S_{e}} \approx C_{H,S_{e}} \equiv C_{S_{e}}$, and $C_{L,S_{x}} \approx C_{H,S_{x}} \equiv C_{S_{x}}$, the HF solution at $\mathbf{z_{n}}$ in the reference domain is reconstructed as:
\begin{align}
    \label{eq:New_HF}
    T_{H,n}^{ref} \, = \, T_{H}^{ref}(\mathbf{z_{n}}) \, &\approx \, T_{H}^{ref}(\gamma) \, C_{T}, \\
    S_{e_{H, n}}^{ref} \, = \, S_{e_{H}}^{ref}(\mathbf{z_{n}}) \, &\approx \, S_{e_{H}}^{ref}(\gamma) \, C_{S_{e}}, \\
    S_{x_{H, n}}^{ref} \, = \, S_{x_{H}}^{ref}(\mathbf{z_{n}}) \, &\approx \, S_{x_{H}}^{ref}(\gamma) \, C_{S_{x}}.
\end{align}
Finally, the reconstructed HF solution in the reference domain is mapped back to the physical domain through the inverse transformation (un-mapping phase), $v_{H}^{ref}(\mathbf{z_{n}}) \, \mapsto \, v_{H}(\mathbf{z_{n}}) \equiv v_{n}(\mathbf{z_{n}})$, yielding the final bi-fidelity prediction for all state variables. The mapping operator and its inverse are presented in detail in the next subsection.

\subsubsection{Bijective mapping framework}
    \label{sec:Mapping_Phase}

This subsection describes the central component of the proposed methodology: the formulation and analysis of the bijective mapping. As discussed above, wildfire propagation is inherently dynamic, with fire fronts evolving as moving combustion waves rather than static spatial structures. Uncertainties arising from model parameters and ambient conditions lead to significant variability in front location, shape, and spread rate, even under seemingly similar scenarios.

In this context, the convection-dominated nature of the governing equations leads to a slowly decaying Kolmogorov $n$-width \citep{Arbes_2025}, implying that hyperbolic-type problems generally require a large number of basis functions to achieve accurate reduced-order approximations \citep{Ohlberger_2016, GreifUrban_2019}. To address this limitation, domain mappings are introduced to align parameter-dependent features within a common reference domain across the training parameter set \citep{Mirhoseini_2023}. By rendering the dominant combustion-front feature stationary in this reference frame--thereby suppressing the dominant convective variability while retaining the remaining diffusive features--the resulting snapshots become effectively compressed and can be accurately represented within a low-dimensional affine subspace. The mapping and inverse mapping procedures for the 1D and 2D formulations are described next, demonstrating the flexibility and robustness of this approach for convection-dominated wildfire dynamics.

\subsection*{One-dimensional front alignment}

For the 1D setting, the mapping operator aligns the dominant combustion-front features of each state variable ($T, S_{e}, S_{x}$) from every snapshot, defined over the spatial coordinate $x \in \R$, onto a common reference location $x_{ref}$ (the midpoint of the spatial domain) and, when necessary, normalizes the effective front width. Let $x \in \big[0, L_{x}\big]$ denote the physical coordinates and $x_{ref} = 0.5 L_{x}$ the prescribed reference point. For a given parameter realization $\mathbf{z} \in I_{z}$, the physical snapshot is $v(\mathbf{z}) = \big[T(\mathbf{z}), S_{e}(\mathbf{z}), S_{x}(\mathbf{z})\big]^{\top}$. The corresponding reference-domain snapshot is obtained through a variable-wise bijective transformation,
\begin{equation}
    \label{eq:Bijection}
    v^{ref}(\mathbf{z})=
    \mathcal{M}\left(v(\mathbf{z}) ; D^{ref}(\mathbf{z})\right),
\end{equation}
where $\mathcal{M}$ denotes the mapping operator and,
\begin{equation}
    \label{eq:1D_Geometry_Descriptors}
    D^{ref}(\mathbf{z}) = \big\{s_T, \, s_{S_e}, \, s_{S_x}, \, \kappa_{S_e}, \, \kappa_{S_x}\big\},
\end{equation}
collects the geometric descriptors of the combustion wave: variable-specific shift parameters ($s_T, s_{S_e}, s_{S_x}$) and stretch factors ($\kappa_{S_e}, \kappa_{S_x}$) that characterize spatial translation and width scaling of the front relative to the reference configuration.

\begin{figure}
    \centering
    \includegraphics[width=.5\textwidth]{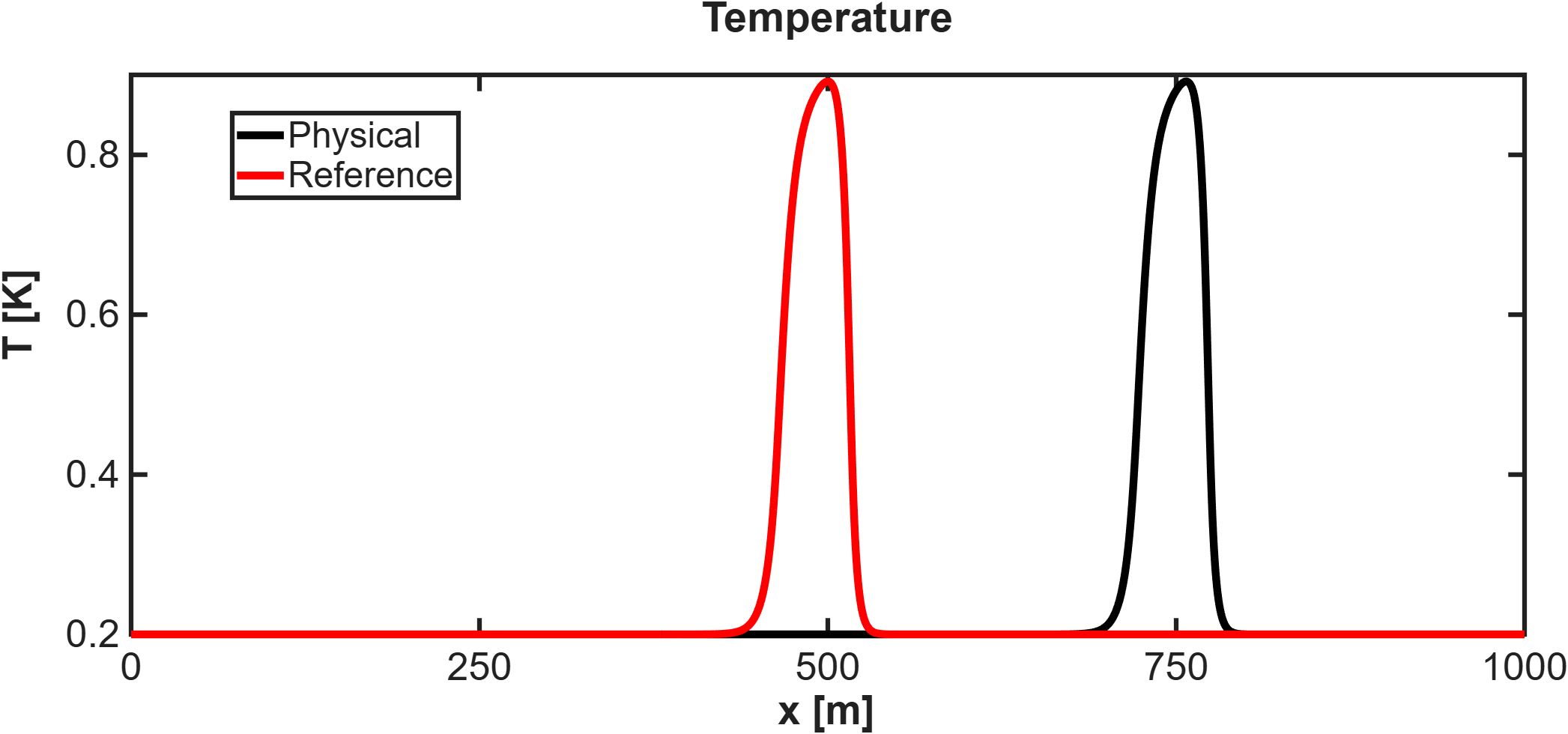} \includegraphics[width=.5\textwidth]{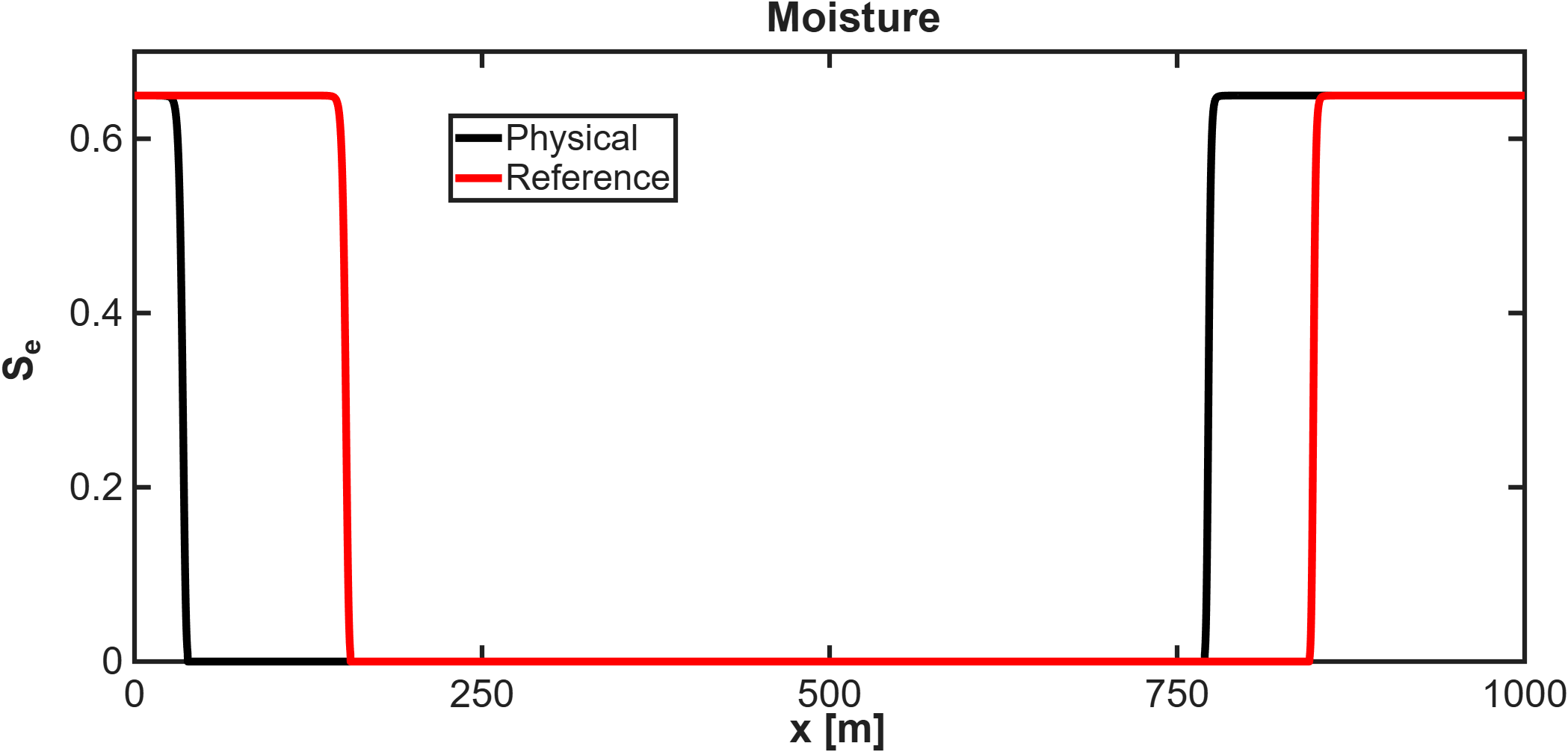} 
    \includegraphics[width=.5\textwidth]{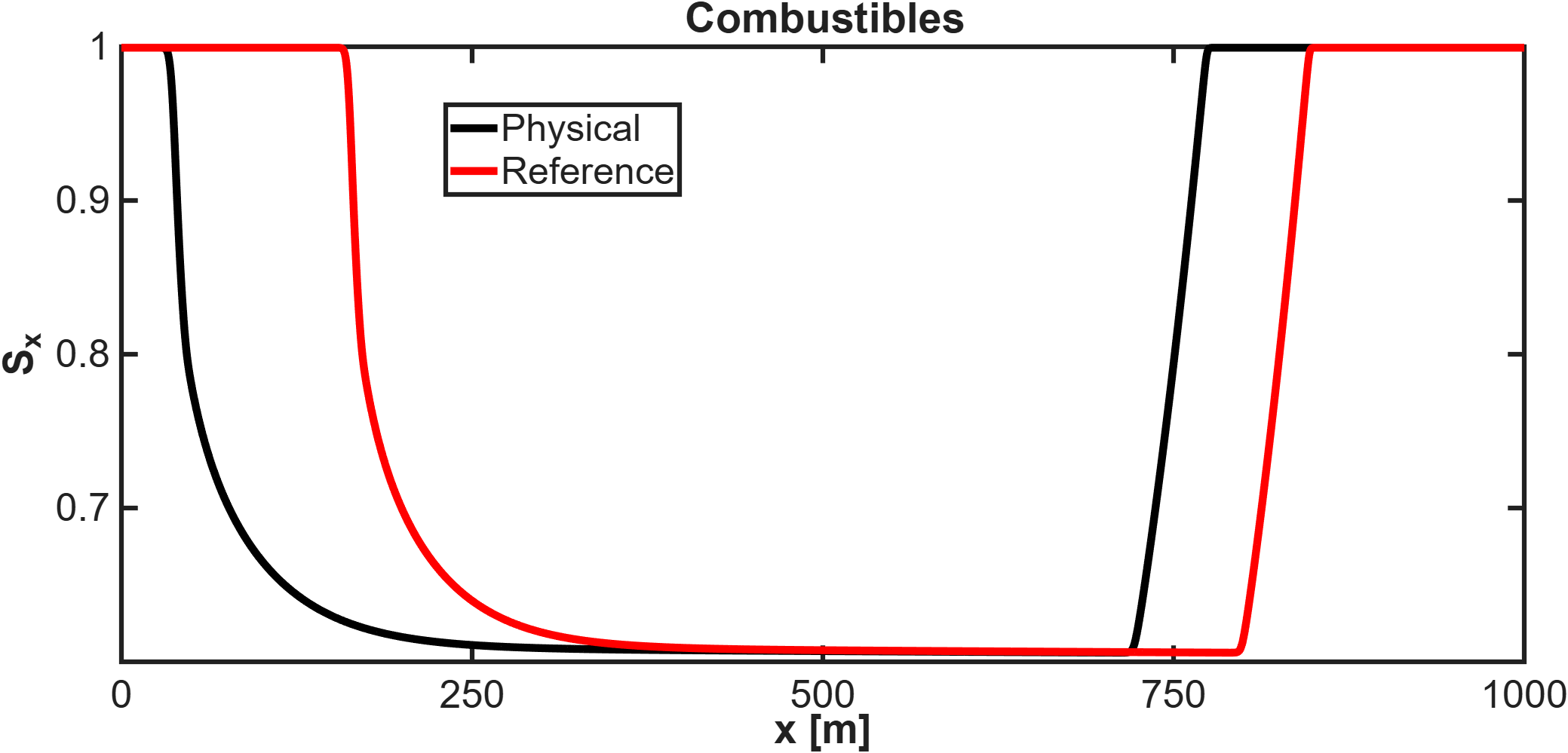} 
    \caption{Mapping of one-dimensional state variables, illustrating snapshots in the physical (black) and reference (red) domains: temperature (shift only), moisture and combustible mass fractions (shift and stretch).}
    \label{fig:Figure_2}
\end{figure}

The temperature distribution exhibits a localized combustion peak (see Fig.~\ref{fig:Figure_2}). Its reference alignment is performed by a pure translation that maps the peak location $x_{T}(\mathbf{z})$ to $x_{ref}$, with the corresponding shift defined as:
\begin{equation}
    \label{eq:1D_T_Shift}
    s_{T}(\mathbf{z}) = x_{ref} - x_{T}(\mathbf{z}).
\end{equation}
The mapped temperature field is then,
\begin{equation}
    \label{eq:1D_Mapped_T}
    T^{ref}(x; \mathbf{z}) = T\left(x - s_{T}(\mathbf{z}); \mathbf{z}\right).
\end{equation}

In contrast, fuel variables $S_{e}$ and $S_{x}$ typically exhibit depletion-induced plateau fronts rather than localized peaks. For each fuel field $S \in \big\{S_{e}, S_{x}\big\}$, two front anchors are extracted from a mid-level set defined by the boundary states (see Fig.~\ref{fig:Figure_2}). Let $y_{L}$ and $y_{R}$ denote representative constant-level values at the left and right boundaries, and define the mid-level value:
\begin{equation}
    \label{eq:Mid_Level_Value}
    y_{mid} = 0.5 \, (y_L + y_R).
\end{equation}
The corresponding left and right crossings $x_L(\mathbf{z})$ and $x_R(\mathbf{z})$ are determined as the locations where $S(x;\mathbf{z})$ intersects $y_{mid}$ on the left (falling) and right (rising) edges, respectively. The resulting physical front width is:
\begin{equation}
    \label{eq:1D_Front_Width}
    w(\mathbf{z}) = x_{R}(\mathbf{z}) - x_{L}(\mathbf{z}).
\end{equation}
Given a prescribed reference width $w_{ref}$ (taken here as the median of the observed widths), the stretch factor $\kappa_{S}(\mathbf{z}) = \{\kappa_{S_{e}}(\mathbf{z}), \kappa_{S_{x}}(\mathbf{z})\}$ is defined by:
\begin{equation}
    \label{eq:1D_Stretch_Factor}
    \kappa_{S}(\mathbf{z}) = \frac{w_{ref}}{w(\mathbf{z})}.
\end{equation}
The shift $s_{S}(\mathbf{z}) = \{s_{S_{e}}(\mathbf{z}), s_{S_{x}}(\mathbf{z})\}$ is chosen so that the mapped left crossing coincides with the reference left edge, $X_{L}$, given by:
\begin{equation}
\label{eq:1D_Left_Edge}
    X_{L} = x_{ref} - 0.5 \,  w_{ref}.
\end{equation}
This yields the affine alignment parameters $(s, \kappa)$, and the mapped fuel field is:
\begin{equation}
    \label{eq:1D_Mapped_S}
    S^{ref}(x;\mathbf{z}) =
    S\left(x_{ref} +
    \frac{x - x_{ref} - s_{S}(\mathbf{z})}{\kappa_{S}(\mathbf{z})}; \mathbf{z}\right),
    \qquad
    S \in \{S_{e}, S_{x}\}.
\end{equation}

The inverse mapping is obtained directly from the forward transformation and is therefore not repeated here. Once a field has been reconstructed in the reference domain, the corresponding geometric descriptors (shift and stretch parameters) are used to recover its physical-domain configuration through the inverse bijection.

In the online phase, the LF geometric descriptors associated with a new realization $\mathbf{z_{n}}$ are computed explicitly from the LF solution. Since the corresponding HF descriptors are unavailable (as no HF simulation is performed), they are inferred via a regression-based procedure. Specifically, using the paired LF and HF geometric parameters evaluated at the selected subset $\gamma$, a low-order polynomial regression model is trained to approximate the LF to HF geometric transfer. Addressing the specific functional form of this transfer, quadratic polynomials are employed to map the spatial shift parameters ($s$), while linear polynomials are utilized for the stretch parameters ($\kappa$). To strictly control error propagation and ensure the physical validity of the surrogate, the predicted $\kappa$ multipliers are mathematically bounded to prevent non-physical front steepening or inversion. Finally, these predicted and bounded HF descriptors are employed in the inverse transformation, thereby restoring the reconstructed HF reference solution to its physical spatial configuration.

\subsection*{Two-dimensional front alignment}

For the 2D formulation, the mapping operator aligns the global combustion footprint of each realization onto a common reference configuration in the spatial domain $(x,y) \in \mathbb{R}^2$. In contrast to the 1D case, where alignment is performed independently per variable through front-specific shift and stretch parameters, the 2D formulation employs a unified geometric description extracted from a non-negative activity indicator that captures the overall fire extent. This unified treatment captures the coupled spatial evolution of the state variables in both coordinate directions, whereas the reduced geometric complexity of the 1D case allows separate variable-wise alignments.

Let the physical snapshot corresponding to $\mathbf{z} \in I_z$ be $v(\mathbf{z}) = \big[T(\mathbf{z}), S_{e}(\mathbf{z}), S_{x}(\mathbf{z})\big]^\top$. The associated reference-domain snapshot is obtained through the bijective transformation in Eq.~(\ref{eq:Bijection}), where $D^{ref}(\mathbf{z})$ now collects the geometric descriptors of the combustion region,
\begin{equation}
    \label{eq:2D_Geometry}
    D^{ref}(\mathbf{z}) = \big\{\mu_x(\mathbf{z}), \, \mu_y(\mathbf{z}), \, \sigma_x(\mathbf{z}), \, \sigma_y(\mathbf{z})\big\}.
\end{equation}
These descriptors are extracted from a non-negative activity indicator $\mathcal{J}(x,y;\mathbf{z})$, defined as:
\begin{equation}
    \label{eq:Activity_Indicator}
    \mathcal{J}(x,y;\mathbf{z}) = \omega \left[T(x,y;\mathbf{z}) - T_{a}\right]^p
    + (1-\omega)\left[1 - S_{x}(x,y;\mathbf{z})\right]^q,
\end{equation}
where $T_{a}$ is the ambient temperature (with $T \geq T_{a}$), $\omega \in (0,1)$, and $p,q > 0$ are prescribed tuning parameters that control the relative contribution of the heating zone (temperature) and the burned region (combustibles), respectively. Temperature thus characterizes active fire behavior, while combustible consumption serves as an indicator of combustion. Moisture evaporation plays a comparatively minor physical role and is excluded in the unified definition. This indicator provides a robust measure of the active combustion footprint and is used exclusively for geometric characterization. Let $M(\mathbf{z})=\iint \mathcal{J}(x,y;\mathbf{z}) \, dx \, dy$ denote the total activity mass. The centroid coordinates (shift) are then:
\begin{equation}
    \label{eq:Centroid_Indicator}
    \mu_x(\mathbf{z}) = \frac{1}{M(\mathbf{z})} \iint x\, \mathcal{J}(x,y;\mathbf{z})\, dx\, dy,
    \qquad
    \mu_y(\mathbf{z}) = \frac{1}{M(\mathbf{z})} \iint y\, \mathcal{J}(x,y;\mathbf{z})\, dx\, dy,
\end{equation}
and the directional spreads (stretch) are:
\begin{equation}
    \label{eq:Spread_Indicator}
    \sigma_{x}(\mathbf{z}) =
    \left(\frac{1}{M(\mathbf{z})}
    \iint (x - \mu_x(\mathbf{z}))^2 \mathcal{J}(x,y;\mathbf{z})\, dx\, dy\right)^{1/2},
    \qquad
    \sigma_y(\mathbf{z}) =
    \left(\frac{1}{M(\mathbf{z})}
    \iint (y - \mu_y(\mathbf{z}))^2 \mathcal{J}(x,y;\mathbf{z})\, dx\, dy\right)^{1/2}.
\end{equation}
The reference statistics $(\mu_{x, ref}, \mu_{y, ref}, \sigma_{x, ref}, \sigma_{y, ref})$ are defined as medians over the training ensemble.

\begin{figure}
    \centering
    \includegraphics[width=.65\textwidth]{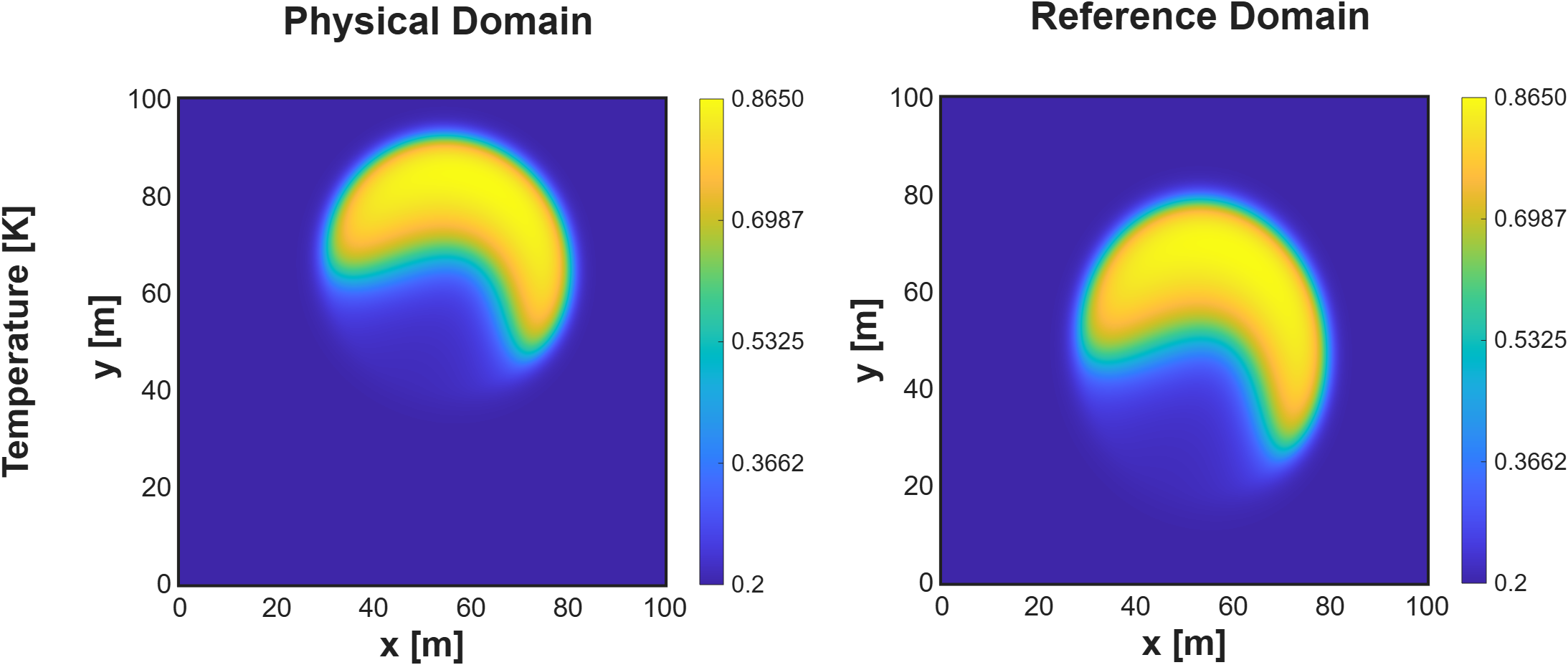} \includegraphics[width=.65\textwidth]{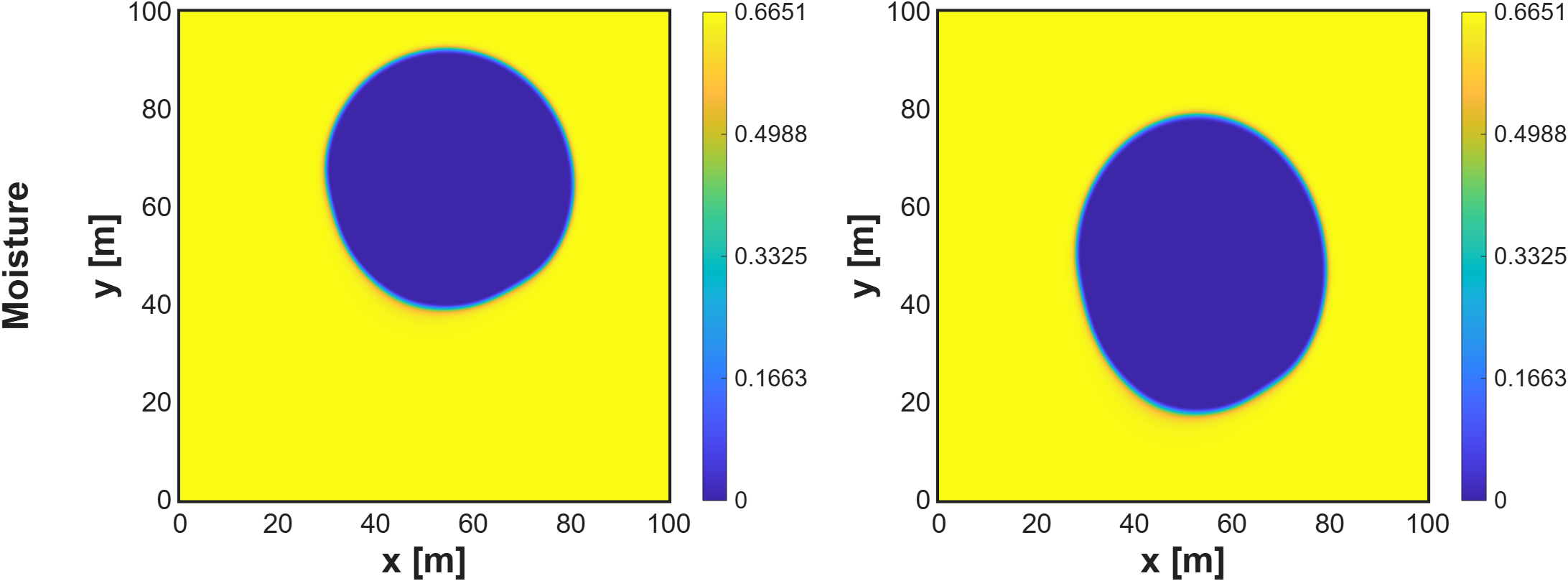} 
    \includegraphics[width=.65\textwidth]{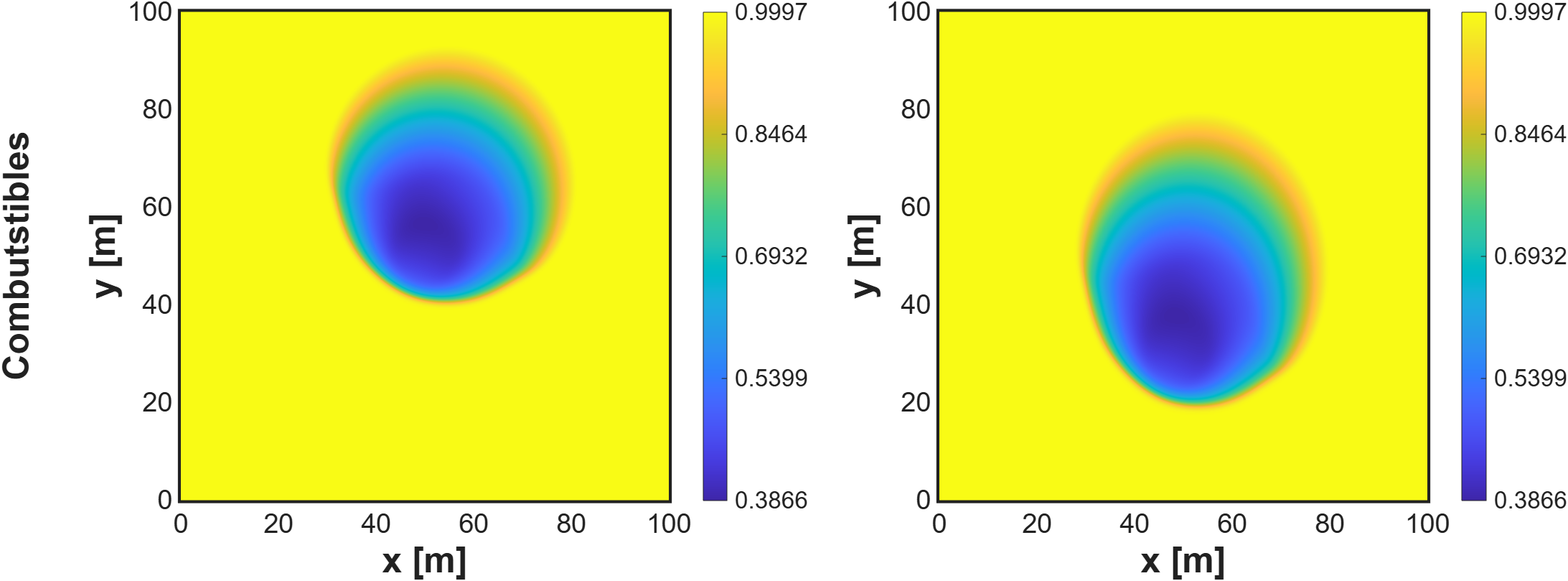} 
    \caption{Mapping of the two-dimensional state variables, showing snapshots in the physical (left column) and reference (right column) domains for temperature, moisture, and combustibles mass fractions. The transformation is driven by a non-negative activity indicator characterizing the heating and combustion region.}
    \label{fig:Figure_3}
\end{figure}

Alignment is then achieved through a separable affine transformation between reference coordinates $(\xi,\eta)$ and physical coordinates $(x,y)$, as denoted below:
\begin{equation}
    \label{eq:2D_Affine_Coordinates}
    x = a_x(\mathbf{z}) \, \xi + b_x(\mathbf{z}),
    \qquad
    y = a_y(\mathbf{z}) \, \eta + b_y(\mathbf{z}),
\end{equation}
with scaling and translation parameters chosen to match centroids and spreads:
\begin{align}
    \label{eq:2D_Scaling_Parameters}
    a_x(\mathbf{z}) = \frac{\sigma_x(\mathbf{z})}{\sigma_{x, ref}},
    \qquad
    b_x(\mathbf{z}) = \mu_x(\mathbf{z}) - a_x(\mathbf{z})\,\mu_{x, ref}, \\
    a_y(\mathbf{z}) = \frac{\sigma_y(\mathbf{z})}{\sigma_{y, ref}},
    \qquad
    b_y(\mathbf{z}) = \mu_y(\mathbf{z}) - a_y(\mathbf{z}) \, \mu_{y, ref}.
\end{align}
The mapped fields are then defined through the pull-back transformation:
\begin{equation}
    \label{eq:2D_Affine_Mapping}
    Z^{ref}(\xi,\eta;\mathbf{z})
    =
    Z\big(a_x(\mathbf{z}) \, \xi + b_x(\mathbf{z}),
    a_y(\mathbf{z}) \, \eta + b_y(\mathbf{z}); \mathbf{z}\big),
    \qquad
    Z \in \{T, S_{e}, S_{x}\}.
\end{equation}
The combustion front is thus translated so that its centroid coincides with the reference centroid and is anisotropically scaled so that its directional spreads match the prescribed reference spreads (see Fig.~\ref{fig:Figure_3}). Unlike the 1D formulation, the same affine transformation is applied to all state variables, reflecting the shared geometric evolution of the multiphysics system. The inverse mapping is obtained directly by inverting the affine transformation in Eq.~(\ref{eq:2D_Affine_Mapping}). After reconstruction in the reference domain, the predicted geometric descriptors are used to recover the physical-domain solution via this inverse bijection.

We note that the affine alignment described above is exact when the front is approximated by an anisotropic Gaussian-like footprint and becomes increasingly approximate as the front develops non-convex or topologically complex shapes (e.g., fingering, breakup, or channeling around fuel heterogeneities). The cases considered in Section~\ref{sec:Results} fall in the regime where the affine description is faithful (extensions to richer mappings are noted as future work).

As in the 1D formulation, the HF geometric descriptors required for the inverse mapping are unavailable during the online stage and must therefore be inferred from their LF counterparts. Accordingly, a functionally analogous but dimensionally expanded regression procedure is employed using the paired LF–HF geometric data from the selected subset $\gamma$. Specifically, linear regressions are used to map the spatial shift parameters ($\mu_x, \mu_y$), while the spatial spread parameters ($\sigma_x, \sigma_y$) are mapped via linear regression strictly in logarithmic space to guarantee positivity. To completely arrest non-physical error propagation during extrapolation, all predicted HF descriptors are mathematically bounded by the absolute extrema (minimum and maximum) of the offline HF training set. These robustly predicted quantities then define the affine transformation needed to restore the reconstructed HF reference solution to its physical 2D spatial domain.

\section{Numerical Results}
    \label{sec:Results} 

This section demonstrates the applicability of the proposed methodology to the developed wildfire spread model in both 1D and 2D domains. For each case, distinct configurations of LF and HF model pairs are established to evaluate the robustness and flexibility of the approach. Section~\ref{sec:1D_Results} explores 1D wildfire expansion (characterized by a linear moving combustion front), where the LF--HF model pair is defined by both mesh refinement (coarse vs.\ fine) and physical complexity (omitted vs.\ full physics). Section~\ref{sec:2D_Results} then examines a more realistic 2D wildfire scenario. Although the LF-HF pair in this case relies solely on mesh-resolution differences, the inherent complexity of the coupled 2D physics still provides a demanding test case. Throughout this section, the HF model serves as the reference ("ground truth") for the LF, conventional bi-fidelity (CF), and mapped bi-fidelity (MF) surrogates. We note that this HF model is itself an interpretable physics-based simulator rather than a detailed CFD or experimental reference, an aspect that is revisited in Section~\ref{sec:Conclusion}.

\subsection{1D Wildfire Case Study}
    \label{sec:1D_Results}

The 1D front propagation, although not constituting a fully realistic physical representation of a wildfire event, provides a fundamental benchmark for demonstrating the novelty of the proposed methodology. The temperature profile exhibits a pronounced peak, whereas the moisture and combustible distributions display smoother, progressively depleted gradients (see Fig.~\ref{fig:Figure_2}). The computational domain is $L_{x}=\big[0, 1000\big]$~m, and the simulation is conducted over the time interval $t_{f} = \big[0, 3600\big]$~s.

Two distinct spatiotemporal discretizations are employed for the LF and HF surrogates. The LF model uses a spatial step of $d_{x_{L}} = 10$~m and a temporal step of $d_{t_{L}} = 0.5$~s, whereas the HF model adopts significantly finer resolutions, $d_{x_{H}} = 0.05$~m and $d_{t_{H}} = 0.01$~s. This disparity affects not only the mesh resolution but also the numerical stability of the underlying ADR system, including CFL constraints and diffusion limits. In addition, the LF surrogate neglects a fundamental physical contribution, exerting a pronounced influence on the morphology of the combustion front. Based on the analytical formulation of the diffusion coefficient in Section~\ref{sec:Model}, the LF submodel excludes the radiation subterm in the diffusion expression--retaining only buoyancy- and wind-shear-induced effects--whereas the HF model incorporates the complete physical formulation. Given that wildfire propagation may be strongly governed by radiative heat transfer (e.g., in plume-driven fire regimes), this distinction carries significant physical implications. Under these two contrasting configurations, a single LF forward simulation requires approximately 4.93~s, compared to 12{,}251.33~s for the HF simulation on the same workstation, highlighting the substantial computational cost of the full-scale model.

In this case study, $M = 4000$ independent samples ($\Gamma \subset I_{z}$) are used for the LF Monte-Carlo simulations, and only $m = 8$ representative collocation points ($\gamma \subset \Gamma$) are selected for the HF simulations. To ensure adequate coverage of the parameter space, LHS is consistently adopted throughout the study. Uncertainty propagation is examined with respect to two key wildfire-related parameters: (i) the ambient wind speed $u_w$ measured at 10~m above the ground and (ii) the initial moisture content $S_{e,0}$. Both parameters introduce significant inherent uncertainty into the model predictions, particularly given the sparsity and limited availability of observational data under real-world circumstances. The sampling domain $I_{z}$ is defined by $u_{w} \in \big[2, 12\big]$~m~s$^{-1}$ and $S_{e,0} \in \big[0.04, 0.16\big]$. Within this domain, the specific values $u_{w}=$ 7~m~s$^{-1}$ and $S_{e,0}=$ 0.1 are considered during the online phase to assess the predictive accuracy of the proposed methodology. The weighting factor associated with the augmented matrix is set to $\beta = 1.0$, while the Tikhonov regularization parameter is $\lambda = 10^{-6}$. Both choices were found to give stable results across a broad range of values. Results are reported for the specified settings to facilitate reproducibility.

\begin{figure}
    \centering
    \includegraphics[width=.78\textwidth]{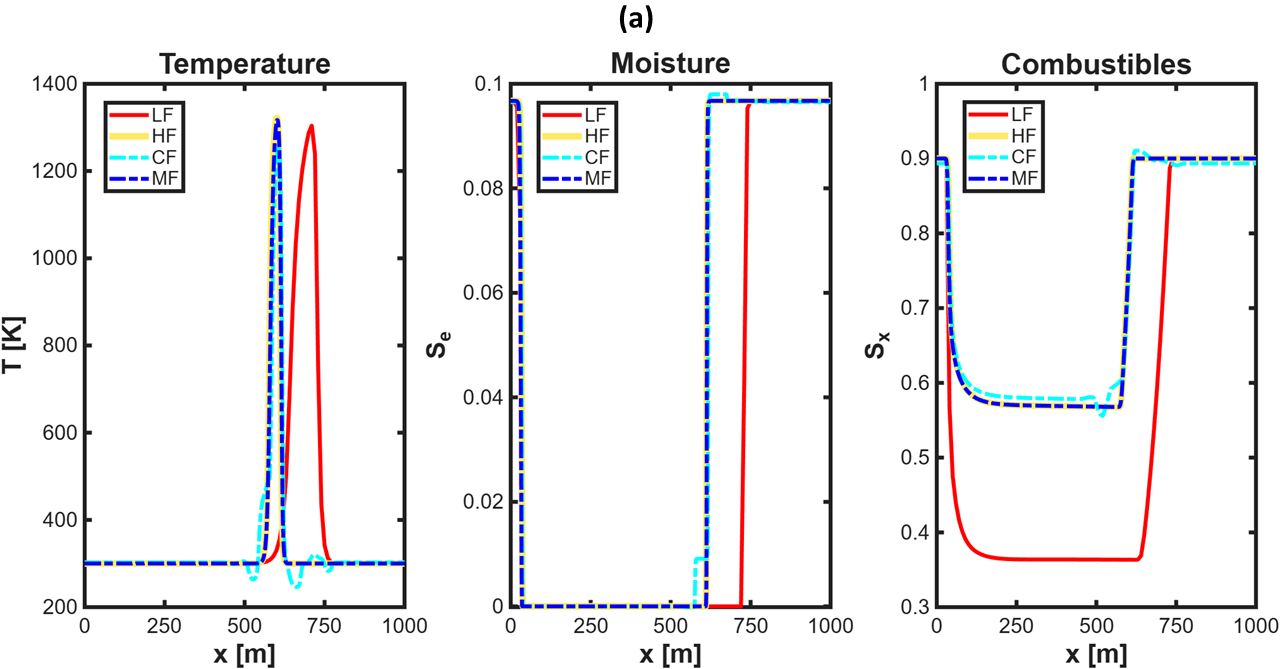}     \includegraphics[width=.76\textwidth]{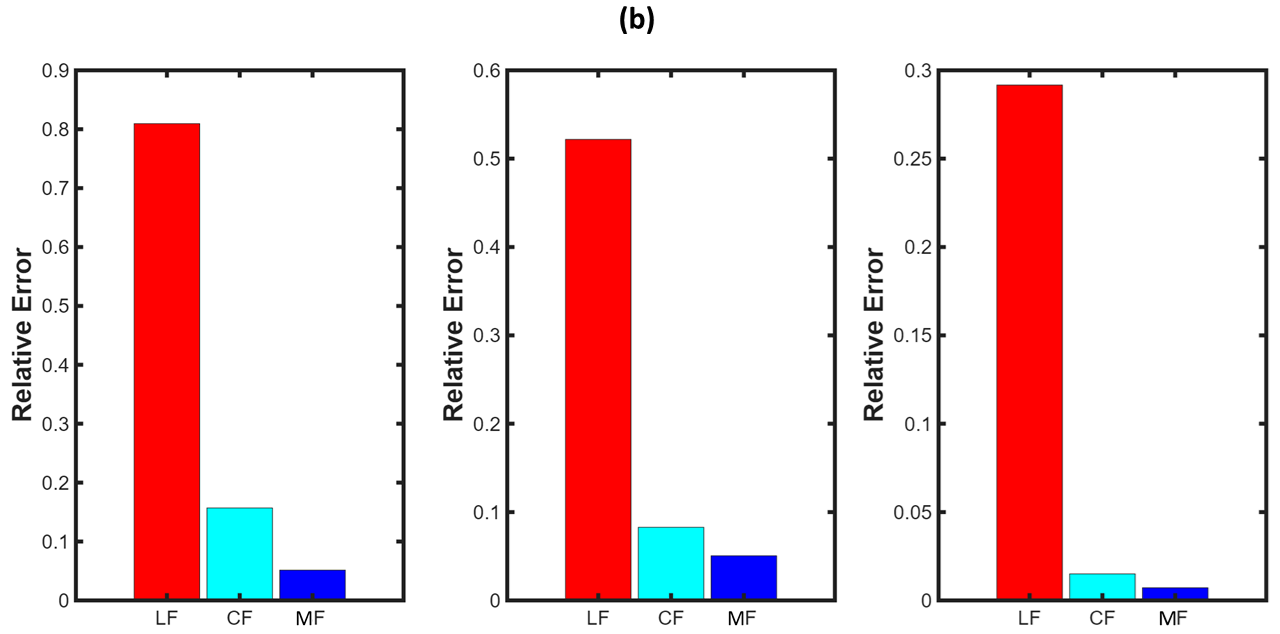} 
    \caption{Comparative analysis of the proposed methodology: (a) model predictions for the three state variables ($T$, $S_{e}$, $S_{x}$), comparing low-fidelity (red), high-fidelity (yellow), conventional bi-fidelity (cyan), and mapped bi-fidelity (blue) responses; (b) relative error bars of the low-fidelity, conventional bi-fidelity, and mapped bi-fidelity models with respect to the high-fidelity reference. In both subfigures, the columns correspond to temperature (first), moisture (second), and combustibles  (third) distributions.}
    \label{fig:Figure_4}
\end{figure}

Figure~\ref{fig:Figure_4}(a) presents the final results obtained using the proposed methodology for the three state variables ($T, S_{e}, S_{x}$). The figure compares the predictions of the nominal HF model, the LF model, and two surrogate bi-fidelity approaches: the conventional bi-fidelity (CF) model, which does not incorporate a bijective mapping between the physical and reference domains \citep{Zhu_2014, Gao_2020, Gao_2021}, and the proposed mapped bi-fidelity (MF) model. The MF approach--by accounting for the convection-dominated nature of the problem--closely reproduces the nominal HF response across all state variables in this 1D setting, aligning their distributed profiles (blue) with the HF solution (yellow). In contrast, the CF model (cyan) exhibits pronounced oscillations in the vicinity of critical regions of the solution (steep gradients) and struggles with the sharp corners of the distributions. These discrepancies arise from the inability of the CF approach to properly represent convection-dominated dynamics, which would otherwise require a substantially larger number of basis modes to capture the shifted distributions accurately. The LF model (red) produces noticeably coarser predictions (phase error), as it neither incorporates the complete set of governing physical mechanisms nor benefits from the fine spatial and temporal discretization of the HF model. Overall, the geometry-aligned transformation in the reference domain effectively eliminates positional inconsistencies, enabling the snapshots to be classified solely according to their intrinsic physical features. The mapping aligns the dominant high-gradient regions of the LF and HF responses, allowing the reduced bases to compare physically corresponding structures rather than spatially shifted fronts. Once the offline training phase is completed, predictions can be performed at a per-query computational cost reduced by approximately three orders of magnitude (essentially requiring only a single LF run) compared to a full HF simulation. Detailed information on computational costs is provided in Table~\ref{table:Computational_Cost} (see Appendix), where the offline training cost is amortized over the number of online queries.

The four approaches are quantitatively evaluated using the relative error to assess their predictive accuracy. Taking the HF response as the reference, the relative error is defined as:
\begin{equation}
    \label{eq:Error}
    \text{Relative Error} =
    \sqrt{\frac{\sum_{i=1}^{N}\big(v_{H,i}(\mathbf{z}) - v_{X,i}(\mathbf{z})\big)^2}{\sum_{i=1}^{N} v_{H,i}(\mathbf{z})^2}},
\end{equation}
where $v_{X}(\mathbf{z})$ denotes the solution obtained from the LF, CF, or MF model and $N$ is the total number of spatial points. As shown in Fig.~\ref{fig:Figure_4}(b), the LF model yields the largest error, while the MF model consistently outperforms the conventional CF approach. These findings underscore the importance of incorporating a bijective mapping when modeling convection-dominated, moving-front phenomena.

\begin{figure}
    \centering
    \includegraphics[width=.85\textwidth]{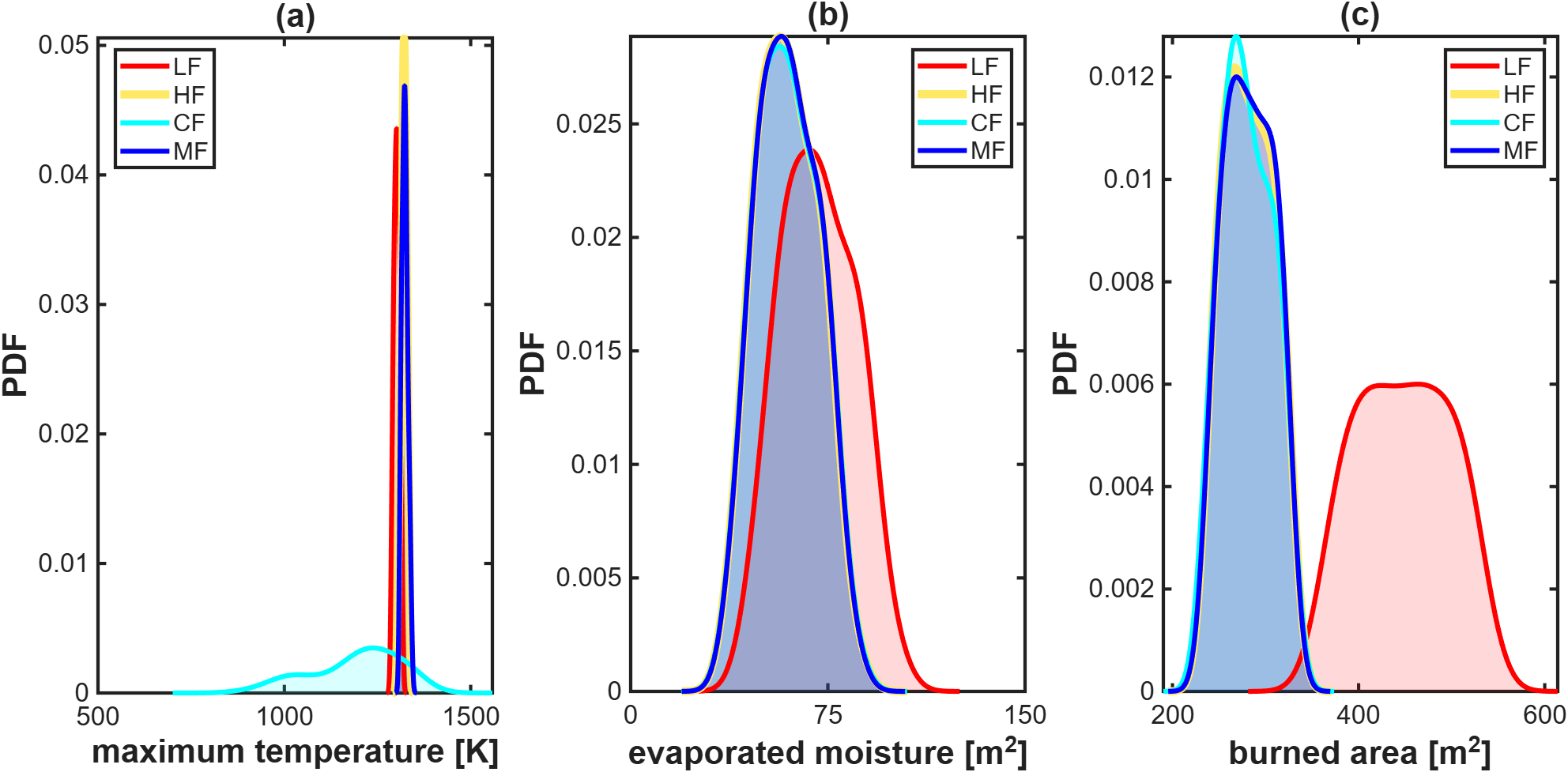} 
    \caption{Probability density functions of the core scalar quantities of interest: (a) maximum temperature, (b) total evaporated moisture, and (c) total burned area. The distributions compare the predictions of the low-fidelity (red), high-fidelity (yellow), conventional bi-fidelity (cyan), and mapped bi-fidelity (blue) models.}
    \label{fig:Figure_5}
\end{figure}

For uncertainty propagation, an MC-based approach is employed to construct the probability density functions (PDFs) of the quantities of interest (QoIs). A set of 50 discrete samples is generated using LHS over the uncertain parameter space defined by $u_{w}$ and $S_{e,0}$. For each sampled realization, the corresponding spatial fields are computed using all fidelity surrogate models (LF, HF, CF, and MF), and a set of scalar QoIs is extracted. The selected QoIs include the maximum temperature ($T_{max}$), the total evaporated moisture ($EM$), and the total burned area ($BA$), capturing fire intensity, moisture evaporation dynamics, and the extent of combustion, respectively:
\begin{equation}
    \label{eq:QoIs}
    T_{max} = \max_{x \in [0, L_{x}]} T(x),
    \qquad
     EM = \int_{0}^{L_{x}} (S_{e,0} - S_e(x)) \, dx,
    \qquad
    BA = \int_{0}^{L_{x}} (1 - S_x(x)) \, dx.
\end{equation}
To obtain continuous probability distributions from the discrete QoI samples, a Gaussian Kernel Density Estimation (KDE) method is applied: a Gaussian kernel is centered at each data point, and the normalized superposition of all kernels yields a smooth approximation of the underlying probability density. We note that the 50-sample reference HF distribution itself carries Monte Carlo error, which is small for the QoIs considered but should be borne in mind when interpreting Fig.~\ref{fig:Figure_5}.

The accuracy of this procedure is illustrated in Fig.~\ref{fig:Figure_5}. Both the CF and MF surrogates reproduce the HF probability density functions with satisfactory agreement for the evaporated moisture and burned area QoIs (see Fig.~\ref{fig:Figure_5}(b, c)). These quantities are cumulative, global spatial integrals, as defined earlier, and are therefore less sensitive to localized discrepancies. As a result, linear combinations, as employed by the CF model, are generally sufficient to approximate such bulk, integrated responses, explaining its ability to match the HF baseline despite its limitations in resolving localized features. In contrast, the CF approach fails completely in predicting the maximum temperature, as shown in Fig.~\ref{fig:Figure_5}(a), producing an excessively flattened and smeared distribution. This failure arises because maximum temperature is highly localized at the moving fire front. Since CF relies on standard linear combinations of spatial basis functions, it lacks the ability to represent translational dynamics. Consequently, when combining snapshots of a propagating front, the method effectively averages spatial positions, leading to non-physical smearing that suppresses peak temperatures and artificially diffuses the thermal field. The mapped bi-fidelity (MF) model overcomes this limitation by incorporating a geometry-aware mapping procedure. By inferring appropriate geometric descriptors, the MF approach realigns the spatial features and restores the statistical distribution to closely match that of the HF model. This demonstrates its accuracy and robustness across both localized quantities, such as maximum temperature, and integrated measures, such as evaporated moisture and burned area. Finally, the LF model, which relies on simplified physics and a coarser spatial resolution, exhibits a pronounced positive bias. In the present scenario, these simplifications lead to an overestimation of fire spread rate and intensity, resulting in significantly inflated predictions of both evaporated moisture and burned area.

\subsection{2D Wildfire Case Study}
    \label{sec:2D_Results}

To further assess the performance of the proposed framework, a 2D case study is conducted to demonstrate its scalability. The complex and interdependent processes governing 2D wildfire propagation render accurate prediction a computationally intensive task. The computational domain is defined by spatial extents $L_{x}, L_{y} \in \big[0, 100\big]$~m, with a final simulation time of $t_{f} = \big[0, 600\big]$~s.

Two distinct grid resolutions are employed to construct the LF and HF models. For the LF configuration, the spatial discretization is $d_{x_{L}}= d_{y_{L}}=$ 2~m with a temporal step of $d_{t_{L}}=$ 0.5~s, while the HF model adopts finer spatial resolutions $d_{x_{H}}=d_{y_{H}}=$ 0.2~m and a temporal step $d_{t_{H}}=$ 0.1~s. This disparity in resolution constitutes the primary distinction between the LF and HF surrogates here, as both models incorporate the complete physical formulation. The associated computational cost differs substantially: a single LF forward simulation requires approximately 3.77~s, whereas the corresponding HF simulation requires 6{,}870.61~s.

To explore the LF sampling domain, $M = 4000$ samples are again generated, and $m = 60$ representative samples are selected via the greedy algorithm of Section~\ref{sec:Offline_Phase} for execution of the HF model. Although this sample size is substantially larger than that employed in the 1D case, it is necessitated by the increased complexity of the multidimensional coupled phenomena. These HF simulations are performed during the offline phase; consequently, in a potential operational or emergent setting, only the computationally efficient online phase of the framework is required. Both environmental conditions and fuel properties are treated as uncertain. Specifically, the ambient wind velocities in the $x$- and $y$-directions are sampled within $u_{w,x}, u_{w,y} \in \big[1, 5\big]$~m~s$^{-1}$, the initial moisture content is defined within $S_{e,0} \in \big[0.04, 0.16\big]$, and the packing ratio (or volume fraction)--a key indicator of fuel compactness--is prescribed within $\alpha \in \big[0.002, 0.008\big]$. While these ranges are employed during the offline training phase, the online prediction stage uses the specific values $u_{w,x}=$ 3.5~m~s$^{-1}$, $u_{w,y}=$ 2.5~m~s$^{-1}$, $S_{e,0}=$ 0.1, and $\alpha=$ 0.005. The algorithmic parameters are set to $\beta = 1.0$ for the augmented matrix and $\lambda = 10^{-6}$ for the Tikhonov regularization. For the activity indicator in the mapping procedure, the parameters are $\omega = 0.85$, $p = 2.0$, and $q = 1.0$. Results were found to be stable to moderate variations in these settings, so these values are kept for the subsequent analysis.

\begin{figure}
    \centering        \includegraphics[width=0.95\textwidth]{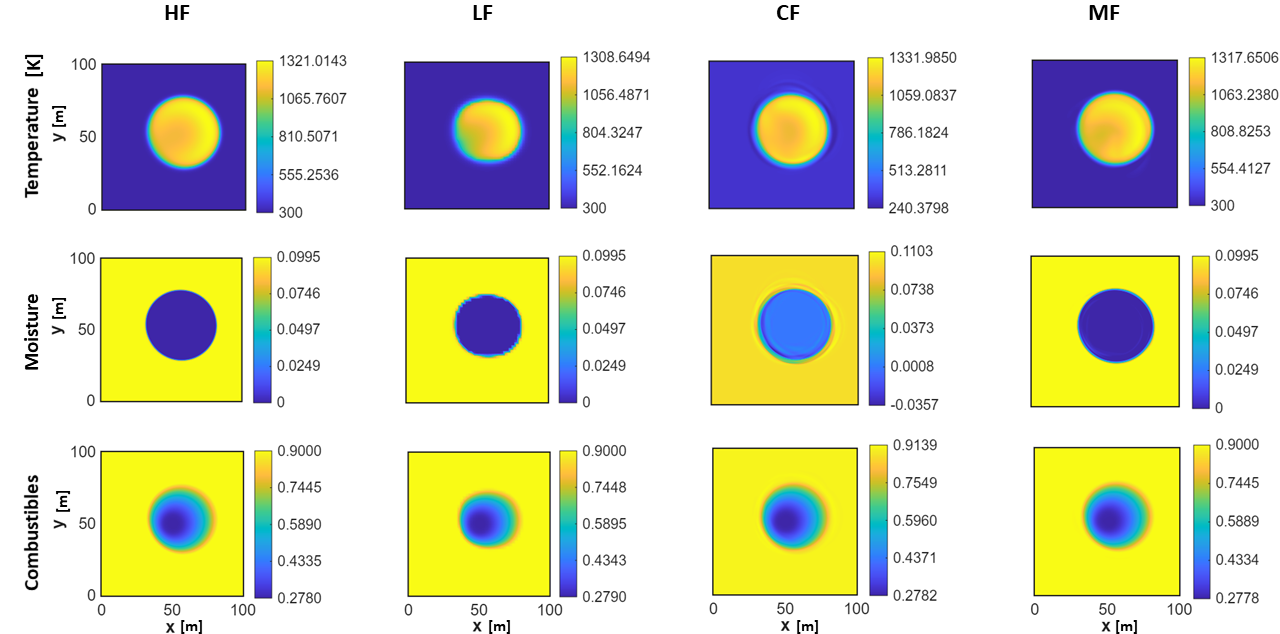} 
    \caption{Comparison of surrogate models. Columns correspond to the high-fidelity (HF), low-fidelity (LF), conventional bi-fidelity (CF), and mapped bi-fidelity (MF) models, respectively, while rows represent temperature (first), moisture (second), and combustibles (third).}
    \label{fig:Figure_6}
\end{figure}

Figure~\ref{fig:Figure_6} presents contour plots of the HF, LF, CF, and MF predictions for each state variable. The MF approach provides the closest visual agreement with the HF solution among the surrogate models. The LF model--due to its coarse resolution--fails to capture the rarefied regions of the fire front associated with smoldering combustion, underestimates the fireline perimeter, and exhibits numerical instabilities. Moreover, it roughly captures the circular shape, but the front is distorted, slightly compressed, and spatially misaligned. The CF model improves over LF but remains insufficient: ambient temperature regions (light blue) are underestimated relative to the HF solution (dark blue), reflecting the limitation of a reduced set of linear basis functions in capturing convection-dominated dynamics. In contrast, the MF prediction closely matches the HF solution, resolving the fire core without spurious oscillations.

A closer inspection of the moisture distributions reveals pronounced Gibbs-type oscillations in the CF model. Oscillatory overshoots near sharp gradients at the fire front persist even with increasing resolution, reflecting the intrinsic limitations of representing discontinuities with smooth basis functions. These oscillations--which produce non-physical negative values--arise from the abrupt transition of moisture to zero, compromising the accuracy of the reduced-order approximation. When the basis functions are aligned in a common reference domain and the bi-fidelity algorithm is applied therein, these oscillations are eliminated. The combustible distribution is captured more smoothly by all surrogate models, with comparable accuracy across approaches.

\begin{figure}
    \centering
    \includegraphics[width=0.78\textwidth]{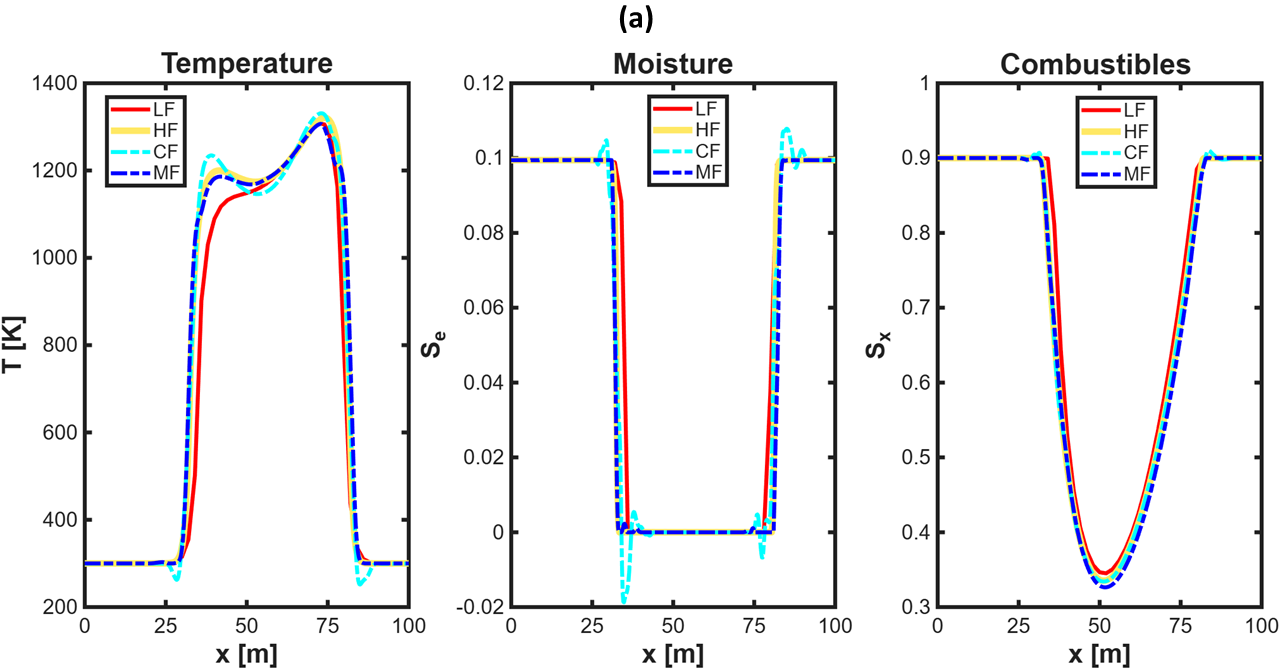}     \includegraphics[width=0.78\textwidth]{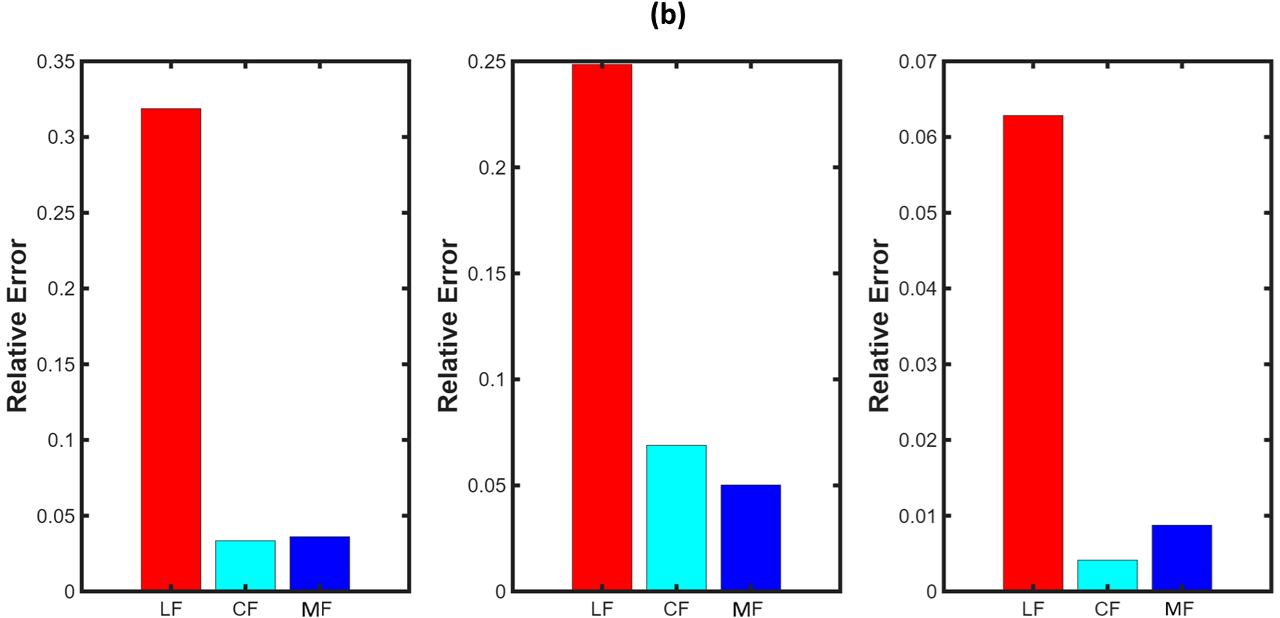}  
    \caption{Comparative analysis of the proposed methodology: (a) model predictions for the three state variables ($T$, $S_e$, $S_x$), comparing low-fidelity (red), high-fidelity (yellow), conventional bi-fidelity (cyan), and mapped bi-fidelity (blue) responses along the section at $y =$ 60 m; (b) relative errors of the low-fidelity, conventional bi-fidelity, and mapped bi-fidelity models with respect to the high-fidelity reference. In both subfigures, the columns correspond to temperature (first), moisture (second), and combustibles (third) distributions.}
    \label{fig:Figure_7}
\end{figure}

To further investigate discrepancies among the surrogate models, a focused analysis is conducted in the fire-core region of the combustion front. Specifically, a horizontal cross-section at $y =$ 60~m is extracted and the corresponding profiles are compared (see Fig.~\ref{fig:Figure_7}(a)). For all state variables, the LF model deviates significantly from the remaining predictions, systematically underestimating the solution. The CF model repeatedly exhibits oscillatory behavior in regions with steep gradients. These differences are quantified through the relative-error metric defined in Eq.~\ref{eq:Error}. As shown in Fig.~\ref{fig:Figure_7}(b), the LF model yields the largest error among all approaches, although the magnitude of the error is reduced compared to the 1D case, since the primary discrepancy between LF and HF stems from the coarser numerical discretization rather than missing physics. For the temperature profile, both the conventional and the mapped bi-fidelity approaches perform similarly. For the fuel-related variables, however, their behavior differs: for the sharp moisture profile, the MF approach yields substantially lower errors, whereas for the smoother combustible profile the CF model is slightly more accurate. This observation suggests that the advantage of the geometry-aligned reconstruction is most pronounced for sharply localized, convection-dominated fronts, while smoother fields may already be well represented by the conventional linear bi-fidelity basis.

Building upon the 1D framework, the uncertainty propagation methodology is extended to a 2D spatial domain. The core post-processing approach of constructing continuous PDFs via Gaussian KDE remains identical, while the underlying MC sampling is expanded to a higher-dimensional uncertainty space. The LHS method for the 2D test set generates 50 unique scenarios across four stochastic input parameters: the ambient wind speed components in both the $x$- and $y$-directions ($u_{w,x}$, $u_{w,y}$), the initial moisture fraction ($S_{e,0}$), and the fuel packing ratio ($\alpha$). The scalar QoIs are then extracted by evaluating the spatial maximum temperature ($T_{max}$) and surface fuel integrals ($EM, BA$) across the 2D grid (Eq.~\ref{eq:QoIs}) to construct the final comparative distributions.

The results indicate that the MF model provides the closest agreement with the HF probability density functions for the selected QoIs, whereas the CF and LF approaches tend to underestimate the corresponding distributions. The deviation of the LF model is expected, as it is based on simplified physical assumptions that limit its predictive capability. Although the CF surrogate attempts to correct these discrepancies, it remains fundamentally limited by its reliance on linear combinations of spatial basis functions. Such linear representations are inherently unable to capture the translational behavior of propagating fronts. As a result, non-physical spatial averaging occurs, leading to artificial smoothing of the solution, with attenuated peak values and a diffused front. Unlike the 1D translating wave where linear combinations completely smeared out the moving peak (for temperature), the 2D fire expands radially from a shared, stationary ignition point. Because all snapshots overlap at this intensely hot central core, the CF model manages to preserve the maximum temperature distribution, even though it fundamentally fails to capture the correct spatial physics along the expanding boundaries. In contrast, the proposed geometry-aligned bi-fidelity (MF) approach overcomes this limitation by employing a spatial bijection to align the dominant physical features prior to coefficient reconstruction. This mapping better accounts for the strongly advective, geometry-dominated variability of the fire and moisture fronts. Consequently, the MF surrogate achieves a level of statistical accuracy comparable to the HF model while maintaining a significantly reduced per-query computational cost.

\begin{figure}
    \centering
    \includegraphics[width=.85\textwidth]{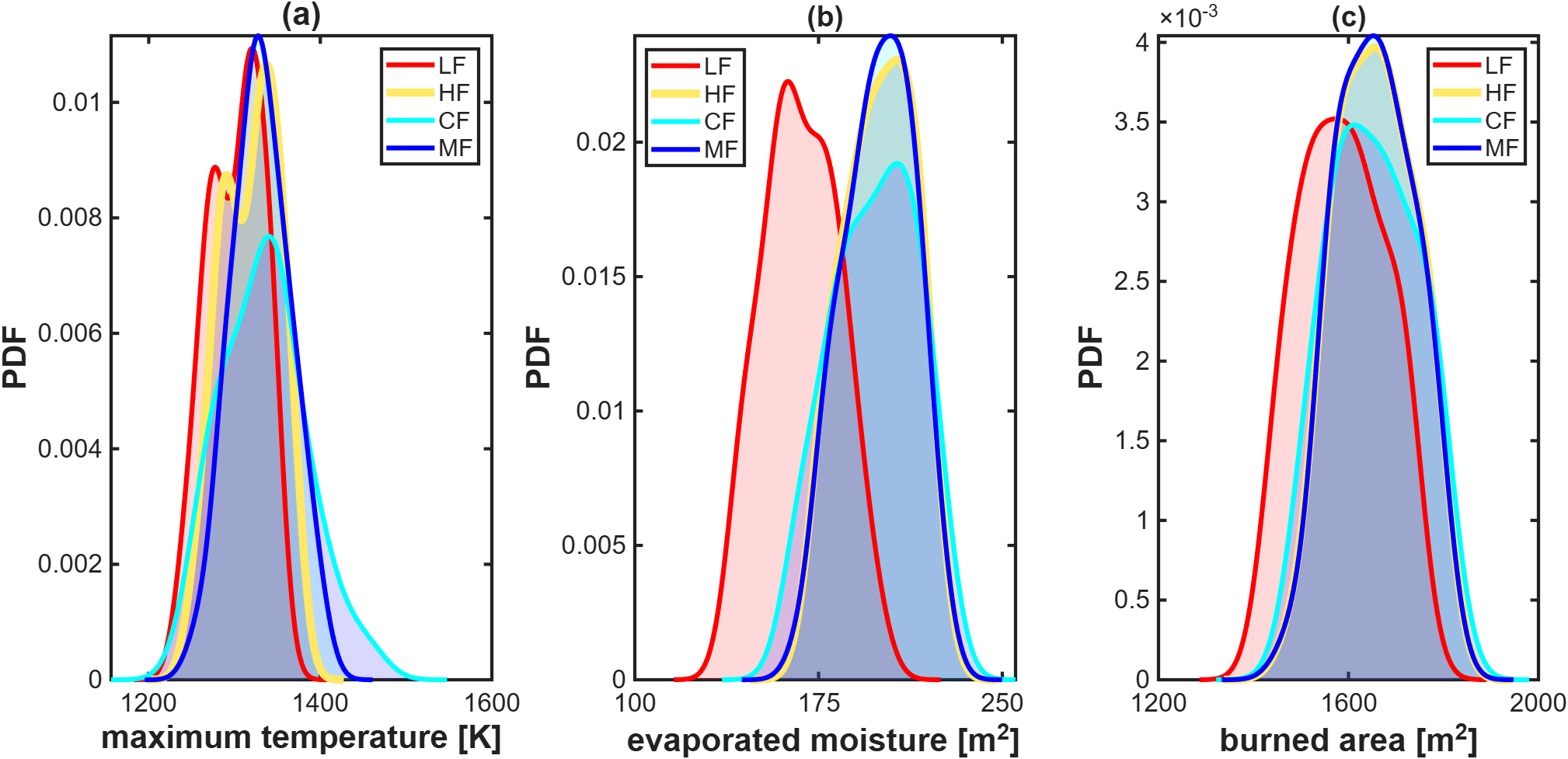}  
    \caption{Probability density functions of the core quantities of interest: (a) maximum temperature, (b) total evaporated moisture, and (c) total burned area. The distributions compare the predictive performance of the low-fidelity (red), high-fidelity (yellow), conventional bi-fidelity (cyan), and mapped bi-fidelity (blue) models.}
    \label{fig:Figure_8}
\end{figure}

\section{Conclusion}
    \label{sec:Conclusion}

This work introduces a geometry-aligned bi-fidelity framework for uncertainty quantification in convection-dominated wildfire spread, demonstrated on the \textsc{ADfiRe} physics-based simulator in both 1D and 2D configurations. The primary contribution is a bijective mapping that transports LF and HF snapshots into a common reference domain prior to basis selection and reconstruction. Rather than directly transferring expansion coefficients between spatially shifted LF and HF snapshots, the proposed approach first maps the dominant front geometry to a common reference frame and then performs basis selection and reconstruction in this aligned space. Building on the general principles of bi-fidelity approximation and multi-fidelity stochastic collocation~\citep{Narayan_2014, Zhu_2014, Gao_2020, Gao_2021}, the method explicitly addresses the moving-front nature of the governing system, separating front geometry from field reconstruction so that reduced bases compare physically corresponding structures rather than displaced ones.

Two case studies were conducted in 1D and 2D configurations. In the 1D case, per-variable shifts and stretches are employed, while in the 2D case, a uniform activity indicator drives a separable affine alignment. Although the 1D case could also be addressed using the activity indicator, the relatively simple underlying physics motivated the adoption of a complementary mapping strategy to provide an alternative perspective on the problem. Across both cases, the LF and HF models differ in physical formulation, mesh resolution, and front position, and the geometry-aligned surrogate yields substantially lower error than the unmapped baseline while suppressing Gibbs-type oscillations near steep gradients. online predictions are roughly three orders of magnitude faster than direct HF evaluation. Accounting for the offline training cost (Table~\ref{table:Computational_Cost}), the framework becomes preferable to direct HF Monte Carlo once the number of online queries exceeds the corresponding break-even point, on the order of a few HF runs in 1D and a few tens of HF runs in 2D.

However, some limitations bound the present results and motivate next steps. First, the HF reference is itself an approximate physics-based model rather than a detailed CFD or experimental measurement. Although the LF-HF gap is meaningful, validation against a more authoritative reference (e.g., FDS/FIRETEC simulations or controlled-burn data) is needed. Second, the affine alignment is most effective for sharply localized, near-convex fronts. Topologically complex fronts (e.g., fingering, breakup, channeling around fuel heterogeneities) likely require non-affine mappings (e.g., optimization-based feature tracking or learned diffeomorphisms). Third, the LF-to-HF transfer of geometric descriptors is currently performed by a low-order polynomial regression, and a more systematic analysis of its error and propagation through to the final fields is warranted. Future work will accordingly target (i) the application of the methodology to real-world wildfire events with appropriate validation data based on ongoing controlled experiments on parallel fire fronts, (ii) extensions of the mapping family to handle non-convex and time-resolved fronts, and (iii) further reductions of the offline cost so that the break-even point relative to direct HF Monte Carlo is shifted earlier, broadening the regime in which the framework is operationally attractive.

\appendix
\section*{Appendix}
    \label{sec:Appendix}

Table~\ref{table:Computational_Cost} summarizes the computational cost of the geometry-aligned bi-fidelity algorithm. For specific steps, the number of CPU cores used is indicated; otherwise a single core is assumed. After completion of the offline training stage, the online bi-fidelity prediction requires only a single LF-level evaluation and is approximately three orders of magnitude faster than a direct HF simulation for the cases considered.

\begin{table}
    \centering
    \caption{Evaluation of the surrogate models' performance across the two case studies, where $M$ and $m$ denote the total numbers of low-fidelity and high-fidelity samples used for offline training. All computational times are reported as wall-clock times (in hours) on a single workstation.}
    \label{table:Computational_Cost}
    \renewcommand{\arraystretch}{1.5}
    \begin{tabular}{ccccc}
    \toprule
    \textbf{Case study: 1D wildfire} & ($M=4000$, $m=8$) & & & \\
    \midrule
    & Offline & & Online & \\
    & Train (LF) & Train (HF) & Test & Total \\
    \midrule
    Low-fidelity (LF) & - & - & $\approx$ \textbf{0.0014} & $\approx$ 0.0014 \\
    High-fidelity (HF) & - & - & $\approx$ \textbf{3.40} & $\approx$ 3.40 \\
    Conventional bi-fidelity (CF) & $\approx$ 0.87 (6 CPU cores) & $\approx$ 8.90 (6 CPU cores) & $\approx$ \textbf{0.0014} & $\approx$ 9.776 \\
    Mapped bi-fidelity (MF) & $\approx$ 0.91 (6 CPU cores) & $\approx$ 10.01 (6 CPU cores) & $\approx$ \textbf{0.0014} & $\approx$ 10.95 \\
    \midrule
    \textbf{Case study: 2D wildfire} & ($M=4000$, $m=60$) & & & \\
    \midrule
    & Offline & & Online & \\
    & Train (LF) & Train (HF) & Test & Total \\
    \midrule
    Low-fidelity (LF) & - & - & $\approx$ \textbf{0.0011} & $\approx$ 0.0011 \\
    High-fidelity (HF) & - & - & $\approx$ \textbf{1.91} & $\approx$ 1.91 \\
    Conventional bi-Fidelity (CF) & $\approx$ 1.41 (6 CPU cores) & $\approx$ 14.57 (6 CPU cores) & $\approx$ \textbf{0.0011} & $\approx$ 16.20 \\
    Mapped bi-Fidelity (MF) & $\approx$ 1.52 (6 CPU cores) & $\approx$ 15.50 (6 CPU cores) & $\approx$ \textbf{0.0011} & $\approx$ 17.32 \\
    \bottomrule
    \end{tabular}
\end{table}

\printcredits

\bibliographystyle{elsarticle-num}

\bibliography{Bibliography}

\end{document}